\documentclass[preprint]{aastex}
\usepackage{epsfig}

\begin{document}

\shorttitle{RM synthesis versus traditional techniques}
\shortauthors{Macquart, Ekers, Feain \& Johnston-Hollitt}

%\title{On the reliability of rotation measure derivation from RM synthesis relative to position angle fitting with application to the Centaurus A field}

\title{On the reliability of polarization estimation using Rotation Measure Synthesis}

\author{J.-P. Macquart\altaffilmark{1}, R.D. Ekers\altaffilmark{1,2}, I. Feain\altaffilmark{2} and M. Johnston-Hollitt\altaffilmark{3}}

\altaffiltext{1}{ICRAR/Curtin University of Technology, Bentley, WA 6845, Australia; J.Macquart@curtin.edu.au}

\altaffiltext{2}{CSIRO Astronomy and Space Science, P.O. Box 76, Epping, NSW 1710, Australia}

\altaffiltext{3}{School of Chemical \& Physical Sciences, Victoria University of Wellington, PO Box 600, Wellington 6140, New Zealand}

\begin{abstract}
We benchmark the reliability of the Rotation Measure (RM) synthesis algorithm using the 1005 Centaurus A field sources of Feain et al.\,(2009).  The RM synthesis solutions are compared with estimates of the polarization parameters using traditional methods.  This analysis provides verification of the reliability of RM synthesis estimates.  We show that estimates of the polarization parameters can be made at lower S/N if the range of RMs is bounded, but reliable estimates of individual sources with unusual RMs require unconstrainted solutions and higher S/N.

We derive from first principles the statistical properties of the polarization amplitude associated with RM synthesis in the presence of noise.  The amplitude distribution depends explicitly on the amplitude of the underlying (intrinsic) polarization signal.  Hence it is necessary to model the underlying polarization signal distribution in order to estimate the reliability and errors in polarization parameter estimates.  We introduce a Bayesian method to derive the distribution of intrinsic amplitudes based on the distribution of measured amplitudes.

The theoretically-derived distribution is compared with the empirical data to provide quantitative estimates of the probability that an RM synthesis solution is correct as a function of S/N.  We provide quantitative estimates of the probability that any given RM synthesis solution is correct as a function of measured polarized amplitude and the intrinsic polarization amplitude compared to the noise.  
\end{abstract}

\section{Introduction}

The birefringence of the magnetized plasma that pervades interstellar and intergalactic space causes  the observed linear polarization properties of most astrophysical radio sources to be strongly frequency dependent.  In cold plasmas the effect of Faraday rotation causes the position angle of the linear polarization vector to increase by an amount $\lambda^2 {\rm RM}$.  The magnitude of the rotation measure (RM) ranges between $\sim 10 \,$rad\,m$^{-2}$ for typical sources observed through the interstellar medium of our Galaxy (Johnston-Hollitt, Hollitt \& Ekers 2004; Schnitzeler 2010) to $\sim 10^3$\,rad\,m$^{-2}$ for the cores of some radio sources to as high as $\sim 5 \times 10^5$\,rad\,m$^{-2}$ for the compact radio source at the centre of the Milky Way, Sgr~A$^*$ (Macquart et al.\,2006, Maronne et al.\,2007).

It is often desirable to derive the rotation measure along a particular line of sight and recover the linear polarization vector intrinsic to the radio source at the point of emission.  This requires observations over a sufficiently large frequency range that the effect of Faraday rotation is measurable but with fine enough spectral resolution that Faraday rotation does not cause appreciable rotation of the polarization vector across individual spectral channels (bandwidth smearing).  However, in many situations of astrophysical interest this trade-off results in a situation where the signal-to-noise ratio of the polarization measurement in each spectral channel is too low to compute either the rotation measure or intrinsic polarization vector without employing a method that simultaneously utilizes the measurements across the entire spectral band.  This is particularly pertinent to observations made with next-generation backends, such as that of the EVLA (Perley et al. 2011) and the Compact Array Broadband Backend (Wilson et al. 2011).

The technique of rotation measure synthesis was developed by (Killeen et al. 1999) to search for polarization from Sgr~A$^*$ in situations where the large expected RM required such high frequency resolution that the signal-to-noise ratio (S/N) per channel was too low. This technique, which was already implicit in the treatment of Burn (1966), was independently discovered by Brentjens \& de Bruyn (2005) to handle the large Galactic rotations seen at low radio frequency.  Over the past few years RM synthesis has emerged as the method of choice for the derivation of Faraday rotation measures from radio polarimetric data.  RM synthesis utilizes both the amplitude and position angle of the polarization vector at all observed frequencies, and is capable of disentangling emission at multiple rotation measures (or Faraday depths) combined within the telescope beam.  Because all the observed frequency dependent information is used without bandwidth smearing and without the nonlinear operations needed to calculate the variation in polarization position angle with frequency, it should also be the optimum method to extract other polarization information from noisy data.  This technique is fundamentally different from previous methods which derived RMs by fitting position angle estimates at a usually small number of often widely spaced frequencies.  

Despite its extensive current use, a number of questions remain concerning the practical application of RM synthesis to real datasets.  This includes debate on the confidence of the results of the technique as a function of S/N (e.g. Law et al.\,2011), and on the reliability of the technique when there are two emission regions with closely-spaced RM and polarization vectors within the same telescope beam (Farnsworth, Rudnick \& Brown 2011).  There is also confusion regarding the appropriate means of treating the effects of polarization bias, associated with the fact that the mean magnitude of the polarization vector is nonzero even if no polarized signal is present (Simmons \& Stewart 1985).  

The objective of this paper is to resolve these issues.  The technique and nomenclature of RM synthesis is summarized in \S\ref{sec:RMsynth}.  In in \S\ref{sec:compare} we proceed to examine several specific facets of the performance of rotation measure synthesis in detail.  We utilize the 1005 source dataset of a field surrounding Centaurus A (Feain et al.\,2009) to examine its performance in estimating  polarization parameters.

In \S\ref{sec:re-analysisAndNoise} we derive the distribution of polarization amplitudes expected from RM synthesis data as a function of the amplitude of the input polarization signal.  We use this to demonstrate the derivation of the intrinsic distribution of polarization amplitudes within a sample.  This is applied to the Cen A dataset to examine the polarization flux density distribution down to the sub-mJy level; this presents a means of measuring the polarization distribution to well below the detection threshold of any given source, in a manner somwhat akin to that of ``stacking'' used in other surveys to achieve statistical detections of populations at detection thresholds below that of an individual source.

In \S\ref{sec:complex} we examine the Cen A dataset for sources with unusual polarization behaviour, including sources with extreme or multiple RMÕs.  Analysis of the changes in the goodness of the fit (as quantified by the reduced-$\chi^2$ statistic) is used to identify the small number of sources with multiple RMs. 
The conclusions are presented in \S\ref{sec:conclude}.

\section{RM synthesis: terminology and summary}\label{sec:RMsynth}

For the purposes of the following sections it is useful to briefly recapitulate how RM synthesis determines the rotation measure and polarization of a source subject to Faraday rotation.  The input data for RM synthesis is a stream of $Q$ and $U$ Stokes parameters, $Q_1, Q_2,\ldots, Q_N$ and $U_1, U_2,\ldots, U_N$ where there are $N$ spectral channels corresponding to wavelengths $\lambda_1, \ldots \lambda_N$.  The polarization data can be equivalently represented as a series of complex numbers where ${\cal P}_j = Q_j + i U_j =  P_j e^{i \psi_j}$.  (Throughout the text we use ${\cal P}$ to denote the complex polarization and $P = |{\cal P}|$ to denote its amplitude.)  RM synthesis works by winding up the measurements of $Q_j$ and $U_j$ for a range of different trial rotation measures, $\phi$.  If the weights corresponding to the spectral channels are denoted $w_j$, then the approximation to the Faraday dispersion function is
\begin{eqnarray}
\tilde F (\phi) &=&  K \sum_{j=1}^N \tilde{\cal P}_j e^{-2 i \phi (\lambda_j^2 -\lambda_0^2)}, \\
\hbox{where } &\null& K = \left( \sum_{j=1}^N w_j \right)^{-1},
\end{eqnarray}
where $\tilde{\cal P}_k = w_k {\cal P}_k$ and it is convenient (but not necessary) to choose $\lambda_0^2$ as the weighted average of the channel values of $\lambda^2$ over the observing band, i.e. $\lambda_0^2 = \sum_{k=1}^N w_k \lambda_k^2 / \sum_{k=1}^N w_k$.  The approximation to the Faraday dispersion function is the convolution of the actual Faraday dispersion function, $F(\phi)$, with a function, $R(\phi)$, which describes the sensitivity of an observation to emission at each Faraday depth, $\phi$: 
\begin{eqnarray}
\tilde{F}(\phi) = F(\phi) \star R(\phi).
\end{eqnarray}
The function $R(\phi)$ is known as the rotation measure transfer function (RMTF) and is computed as follows:
\begin{eqnarray}
R(\phi) = K \sum_{j=1}^N w_j e^{-2 i \phi (\lambda_j^2 - \lambda_0^2)}.
\end{eqnarray}
In the simple case in which all channel weights are equal, the above relations simplify to 
\begin{eqnarray}
\tilde F (\phi) &=&  \frac{1}{N} \sum_{j=1}^N {\cal P}_j e^{-2 i \phi (\lambda_j^2 -\lambda_0^2)}, \label{Fdefn} \\
R(\phi) &=& \frac{1}{N} \sum_{j=1}^N e^{-2 i \phi (\lambda_j^2 - \lambda_0^2)}.
\end{eqnarray}
Now suppose a source contains polarization emission at one or more Faraday depths, $\phi_k$, and associated with each depth is a polarization vector ${\cal P}_{\rm src,k}$.
It is clear from eq.\,(\ref{Fdefn}) that for a dataset of polarization measurements for a source with Faraday depths $\phi_0$, $\phi_1,\ldots$ the sum over channels results in a coherent summing of the polarization vectors, and the amplitude of the function $F(\phi)$ will be large.  The tendency for this sum to add coherently for values of $\phi_k$ that match (i.e. correctly de-rotate) the polarization data means that $|F(\phi)|$ peaks at the locations of the rotation measure(s) in the source, with the quantity $|F(\phi_k)| =  |{\cal P}_{\rm src,k}|$ giving the amplitude of the polarization vector that is rotating at each Faraday depth, $\phi_k$.  Thus the problem of measuring the Faraday depths and rotation polarization vectors that are significant in a given dataset is one of locating those peaks in the function $|\tilde F(\phi)|$ that exceed some given threshold.

\subsection{RM synthesis deconvolution implementation (CLEAN)}
% Clean:
It is often necessary to recover the Faraday dispersion function, $F$, by deconvolving $\tilde F$ from the RMTF, particularly if the source contains emission at multiple Faraday depths and if the S/N of the observations is high\footnote{When the S/N is sufficient, this enables us to ``super-resolve'' $\tilde F$ and search for multiple peaks in $\phi$ a scale smaller than the width of $R$.}.  To this end, we implemented a variation of the  CLEAN algorithm (H\"ogbom 1974; Heald, Braun \& Edmonds 2009) to handle the fact that both $\tilde F$ and the RMTF are complex-valued.  It searches for the highest peak in $|\tilde F|$ over a specified range of $\phi$ and removes a suitably scaled copy of the RMTF centred on its corresponding Faraday depth.  The algorithm iterated over successive peaks until either a maximum number of allowed iterations was reached or the rms of the fluctuations in $\tilde F$ in the search region matched the rms of $\tilde F$ exterior to the search region.  We used a loop gain of $\gamma=1.0$ in the CLEAN algorithm.  Lower loop gain values, as normally used in 2D image deconvolution, were trialled but it was found that since these stop recovering the polarization signal once it reaches the noise floor; solutions in which $\gamma < 1$ are used systematically underestimate the polarization amplitude by at least $1 \sigma$.  

% Output
The RM synthesis code was supplemented by several additional features.   Because the functions $\tilde F$ and $R$ were computed on a grid of Faraday depths much finer than the width of $R$, it is sometimes desirable to group together CLEAN components that are very tightly clustered together and identify them as members of a single component.  By grouping components in this way we could obtain the $\chi^2$ estimate for each independent component.  We introduced a Faraday depth bunching tolerance, $\Delta \phi$, such that any two components with depths $\phi_j$ and $\phi_k$ were merged if $| \phi_j - \phi_k | \leq \Delta \phi$.   A tolerance $\Delta \phi = 5\,$rad\,m$^{-2}$ was found to be suitable for the dataset examined in the present paper.

The code output a list of clean components ranked in order of amplitude, together with their corresponding Faraday depth and polarization vector ${\cal P}_i$ evaluated at the mean square wavelength $\lambda_0$.  The code also computed the reduced $\chi^2$ associated with the polarization solution as a function of the number of clean components used.   To be specific, for measurements with standard deviations $\sigma_Q$ and $\sigma_U$ in $Q$ and $U$ respectively, and clean components with Faraday depths ${\phi_{c}}_1, {\phi_{c}}_2, \ldots$ and associated polarization vectors ${{\cal P}_c}_1,{{\cal P}_c}_2, \ldots$, the reduced $\chi^2$ associated with a fit to the data by including the first $n$ clean components is,
\begin{eqnarray}
\chi^2 = \frac{1}{2 N - 3 n} \sum_i^N \frac{[Q_i - Q({\phi_c}_1,\ldots, {\phi_c}_n;{{\cal P}_c}_1,\ldots,{{\cal P}_c}_n;\lambda_i) ]^2}{\sigma_Q^2} + 
	\frac{[U_i - U({\phi_c}_1,\ldots, {\phi_c}_n;{{\cal P}_c}_1,\ldots,{{\cal P}_c}_n;\lambda_i) ]^2}{\sigma_{U}^2},  \label{chi-square}
\end{eqnarray}
where 
\begin{eqnarray}
Q({\phi_c}_1,\ldots, {\phi_c}_n;{{\cal P}_c}_1,\ldots,{{\cal P}_c}_n;\lambda_k)  + i U({\phi_c}_1,\ldots, {\phi_c}_n;{{\cal P}_c}_1,\ldots,{{\cal P}_c}_n;\lambda_k)   = \sum_{j=1}^n {{\cal P}_c}_j e^{2 i \lambda_k^2 {\phi_c}_j}.
\end{eqnarray}
This $\chi^2$ statistic was found to be an excellent estimator of the number of significant clean components required to model the polarization behaviour of each source.  It was also useful to identify cases in which RM synthesis solution differed significantly from the polarization behaviour of the source; this highlighted cases in which the polarization is not noise-like but where its behaviour is chiefly governed by a mechanism other than Faraday rotation (\S\ref{sec:complex} discusses specific examples of this).

\section{RM synthesis of Centaurus A data}\label{sec:compare}

In this section we compare the performance of RM synthesis against other techniques of determining the polarization and rotation measures of radio sources.
As an initial step, we investigate the accuracy of the rotation measures reported by RM synthesis as a function of signal-to-noise ratio.  
We then compare the rotation measures derived by RM synthesis against a technique that mimics the technique used to derive the rotation measures extracted from the NRAO VLA Sky Survey (Condon et al.\,1998; Taylor, Stil \& Sunstrum 2009) based on measurements of the polarization in two closely separated bands at 1.4\,GHz.  We then compare the performance of RM synthesis against a more sophisticated technique that performs a least-squares fit to the polarization data.  

This analysis is based on the ATCA observations of 1005 background sources contained
in a $\sim34\,$deg$^{2}$ field centred on the radio galaxy Centaurus A (Feain et al.\,2009). The data have a bandwidth of 192 MHz divided into 24 $\times$ 8 MHz channels between 1296 and 1480 MHz.   Because of the effect of the reweight option in Miriad a 5\% correlation was introduced between channels resulting in 22.75 independent channels.  
The total bandwidth and spectral resolution of these data mean that the FWHM in Faraday rotation for a single polarization component is 280 rad m$^{-2}$, and that it is sensitive to RMs with magnitudes less than $\approx$ 3500 rad m$^{-2}$.  The relatively low RM resolution in these data avoids the potential complexity in the interpretation of any sources with closely-spaced multiple RMs, as discussed by Farnsworth, Rudnick \& Brown (2011).  We allowed our RM synthesis code to search to RMs of $\pm 4000\,$rad\,m$^{-2}$.   The average root-mean-square error in U and Q is 0.089\,mJy and the corresponding average sigma in $P$ (which is the measure of noise used throughout this paper) is 0.12\,mJy.

\subsection{Confidence versus S/N based on RM synthesis-derived parameters} \label{subsec:StoN}
We performed RM synthesis on the 1005 sources in the Cen A field to investigate the reliability of the derived Faraday depths as a function of S/N ratio.  Several recent papers adopt a S/N of $7$ as the threshold below which the results of RM synthesis are regarded as unreliable and are not reported (Feain et al.\,2009; Law et al.\,2011).   It is therefore propitious to investigate whether this threshold represents a sharp barrier to the believability of RM synthesis results, or whether results below this level only decline gradually in accuracy.  The large size of our sample enables us to empirically derive the level of confidence associated with RM synthesis solutions over a large range of S/N ratios.

\begin{figure}[ht]
\includegraphics[angle=0,width=0.9\textwidth]{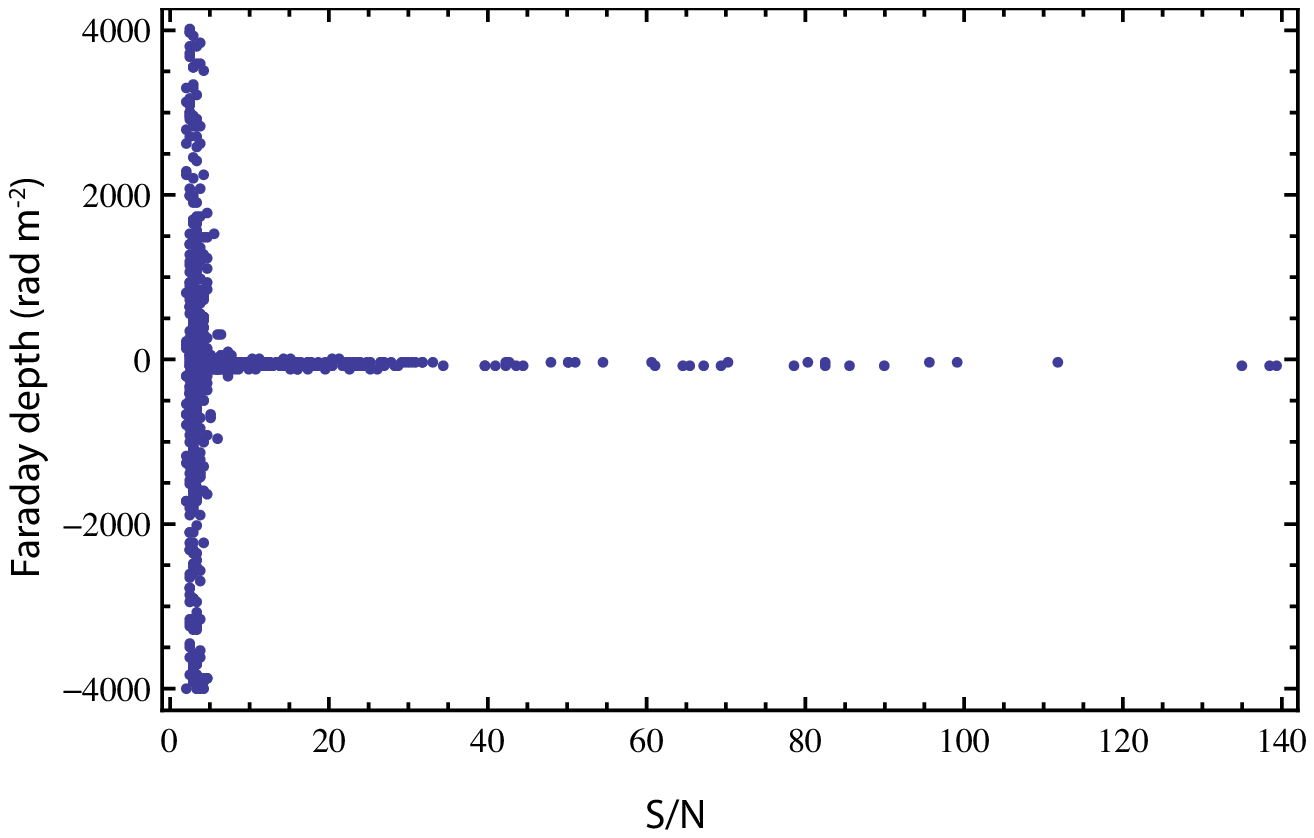}
\includegraphics[angle=0,width=0.9\textwidth]{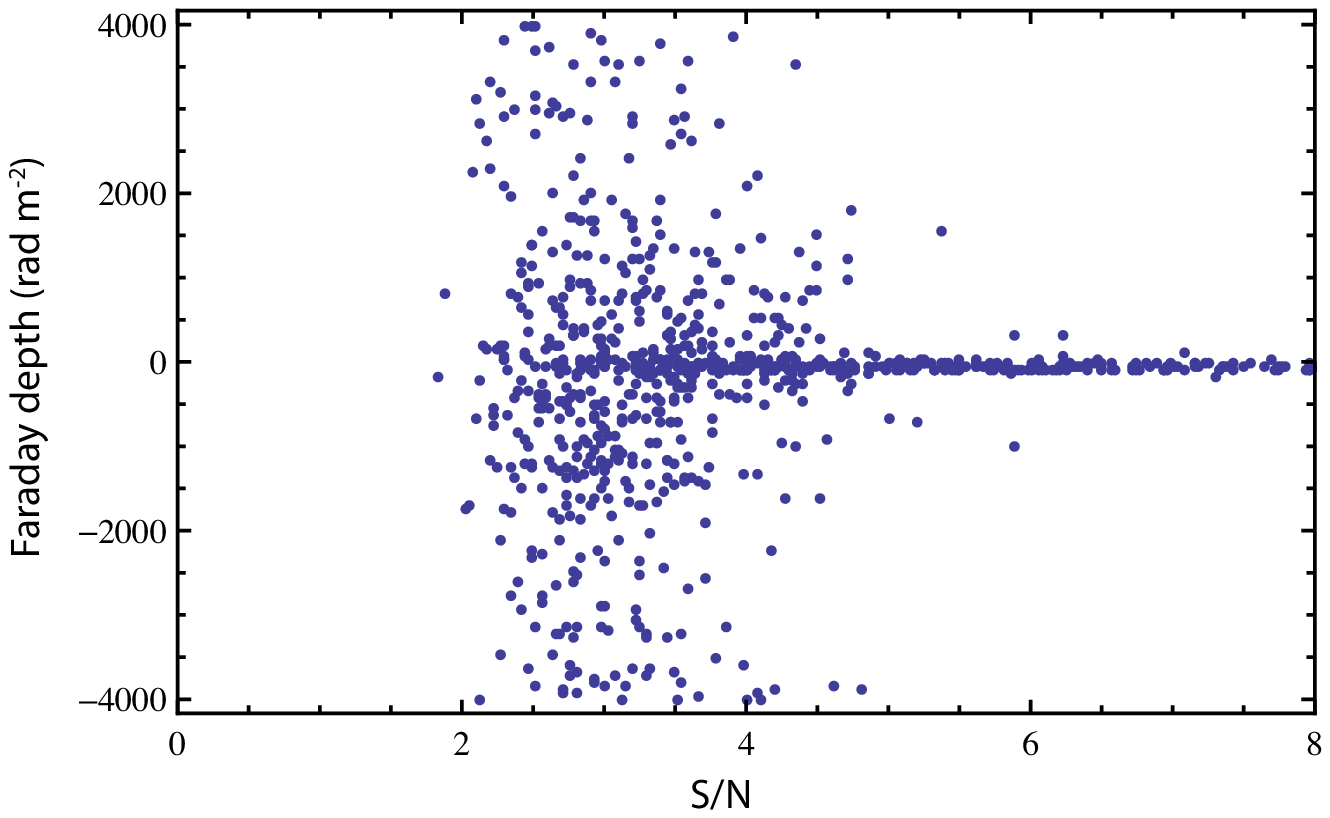}
\caption{Scatter plots showing the distribution of Faraday depths as a function of the signal-to-noise ratio of the polarization detection.  The top panel shows the entire dataset while the
lower panel provides a more detailed plot for S/N$<8$.} \label{fig:SNscat}
\end{figure}

\begin{figure}[ht]
\begin{center}
\includegraphics[angle=0,width=0.6\textwidth]{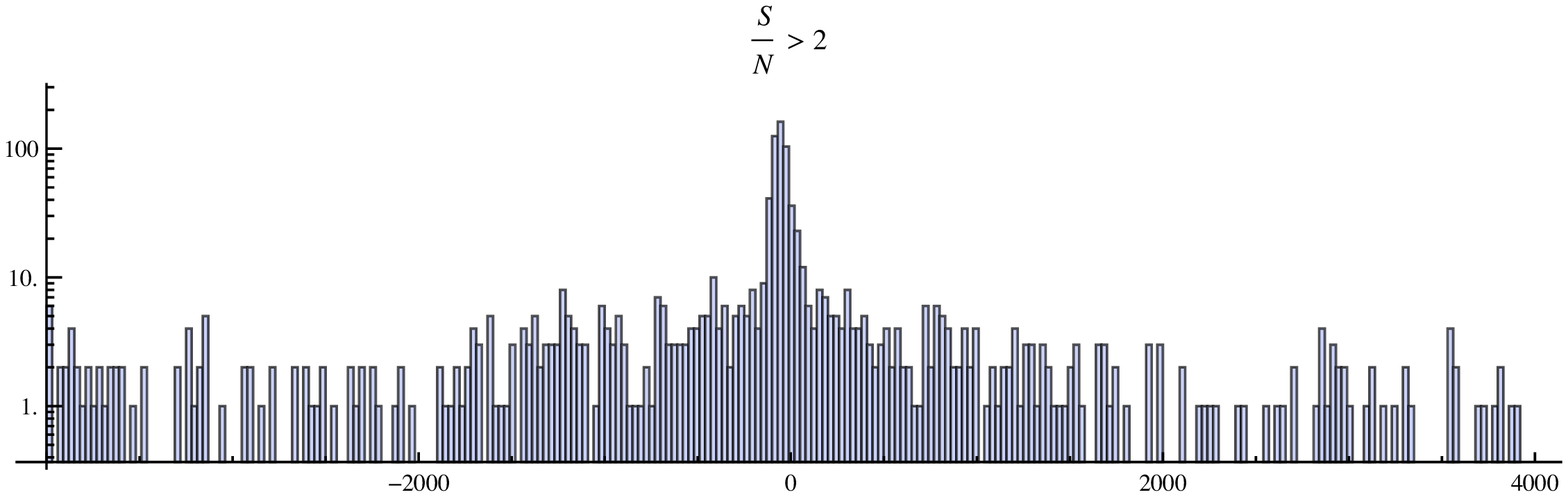}
\includegraphics[angle=0,width=0.6\textwidth]{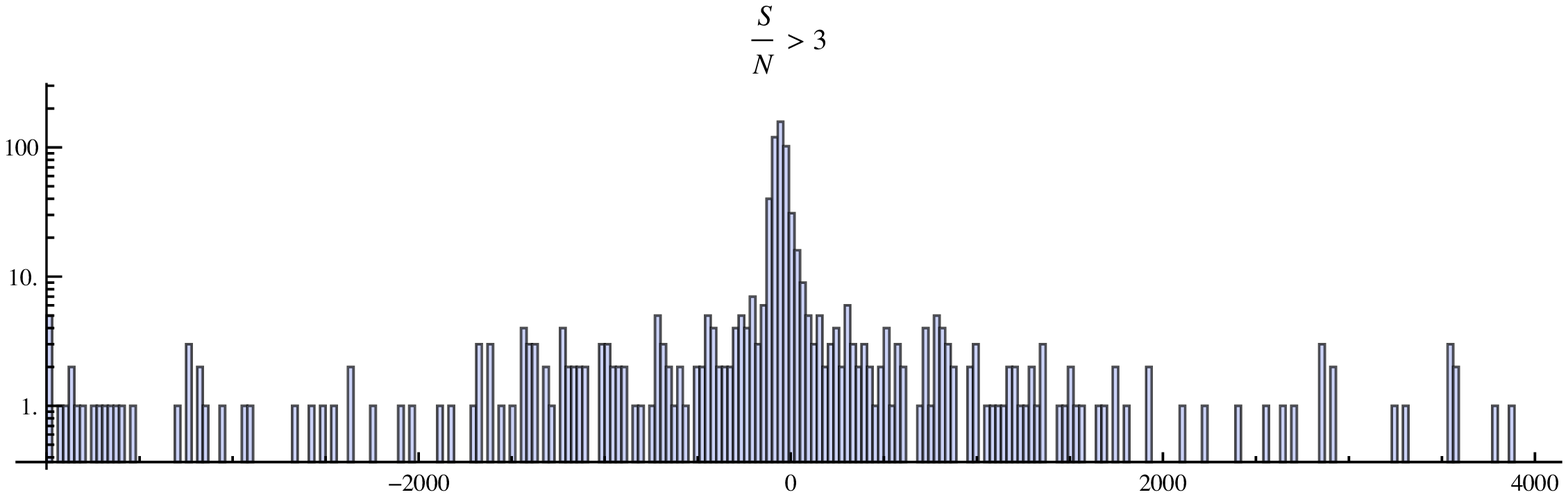}
\includegraphics[angle=0,width=0.6\textwidth]{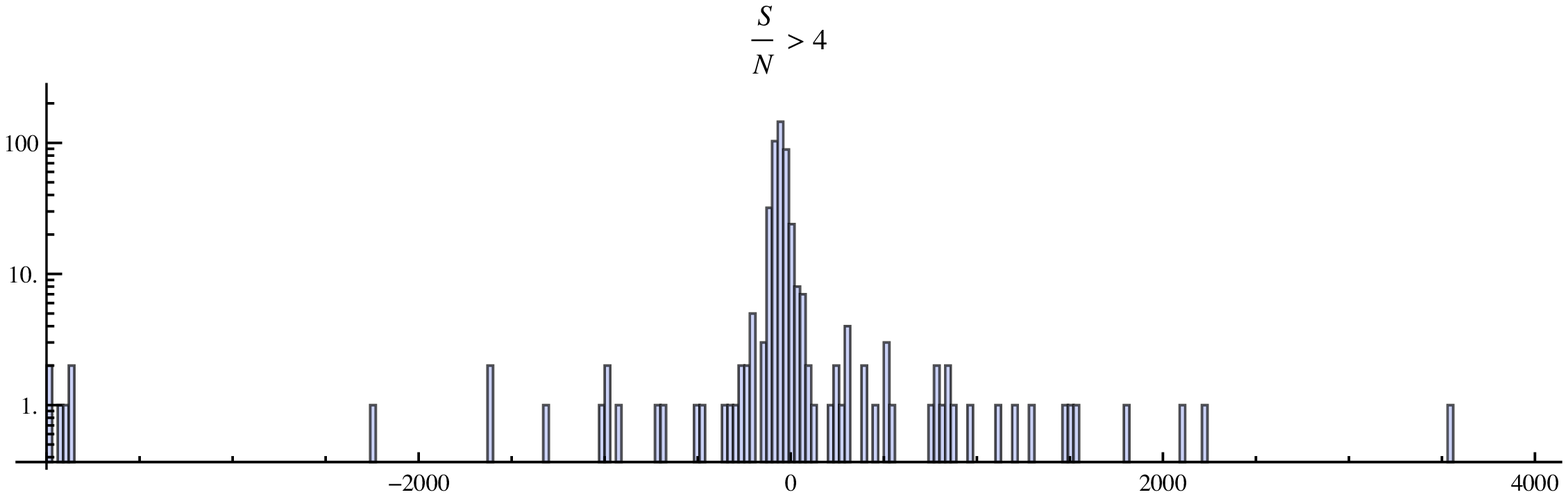}
\includegraphics[angle=0,width=0.6\textwidth]{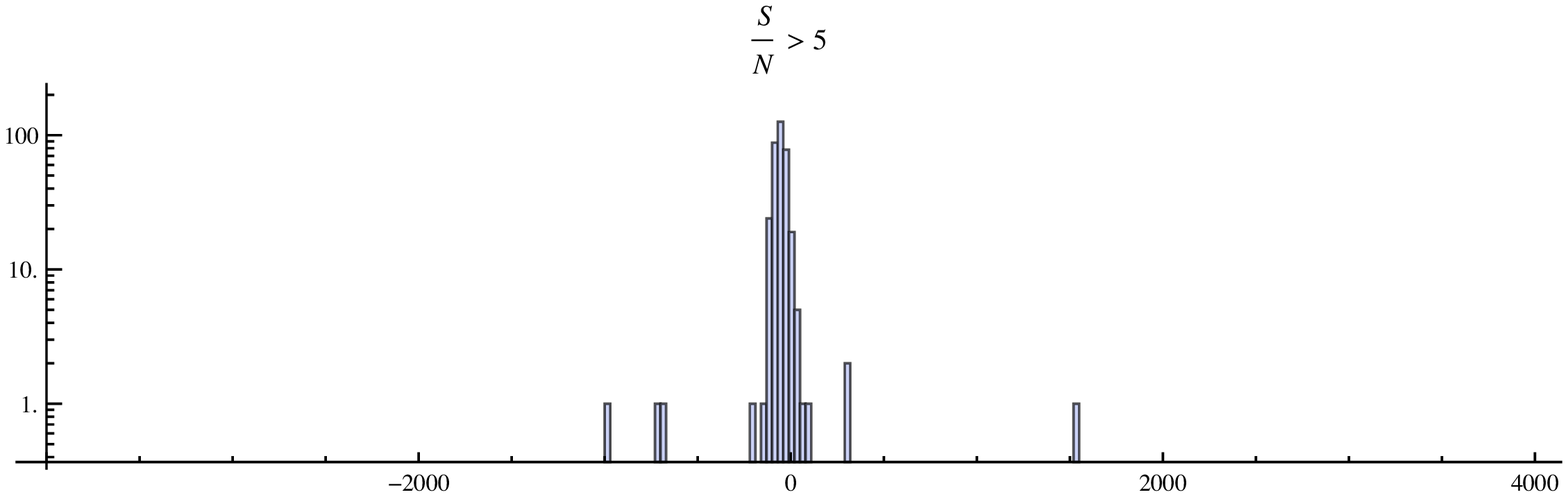}
\includegraphics[angle=0,width=0.6\textwidth]{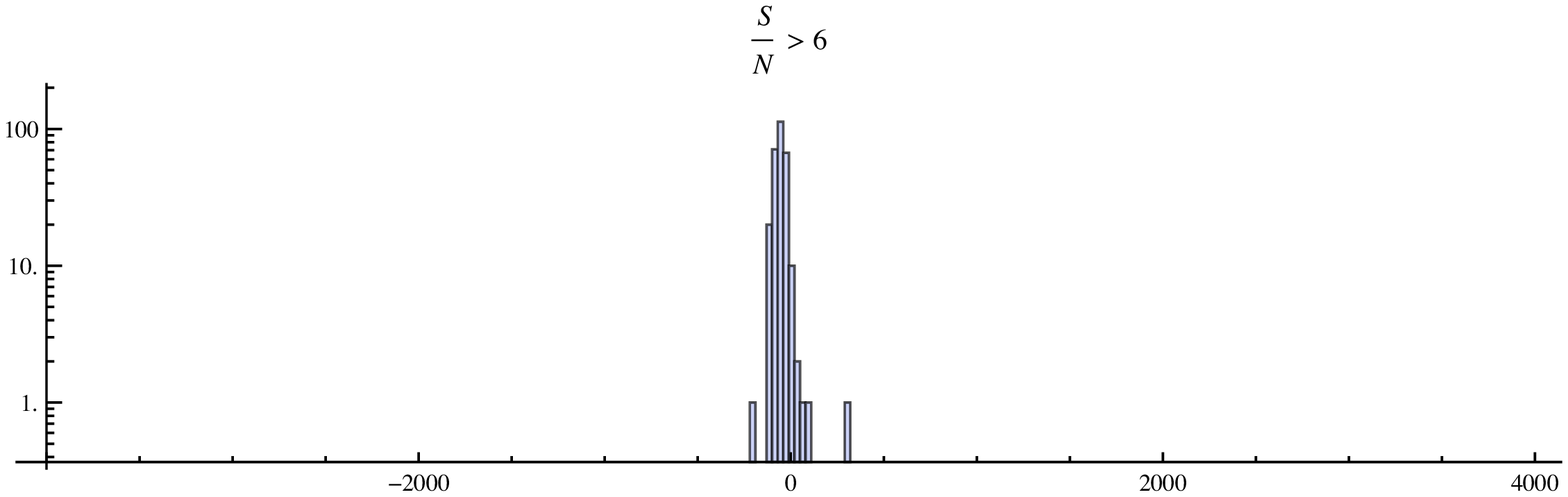}
\includegraphics[angle=0,width=0.6\textwidth]{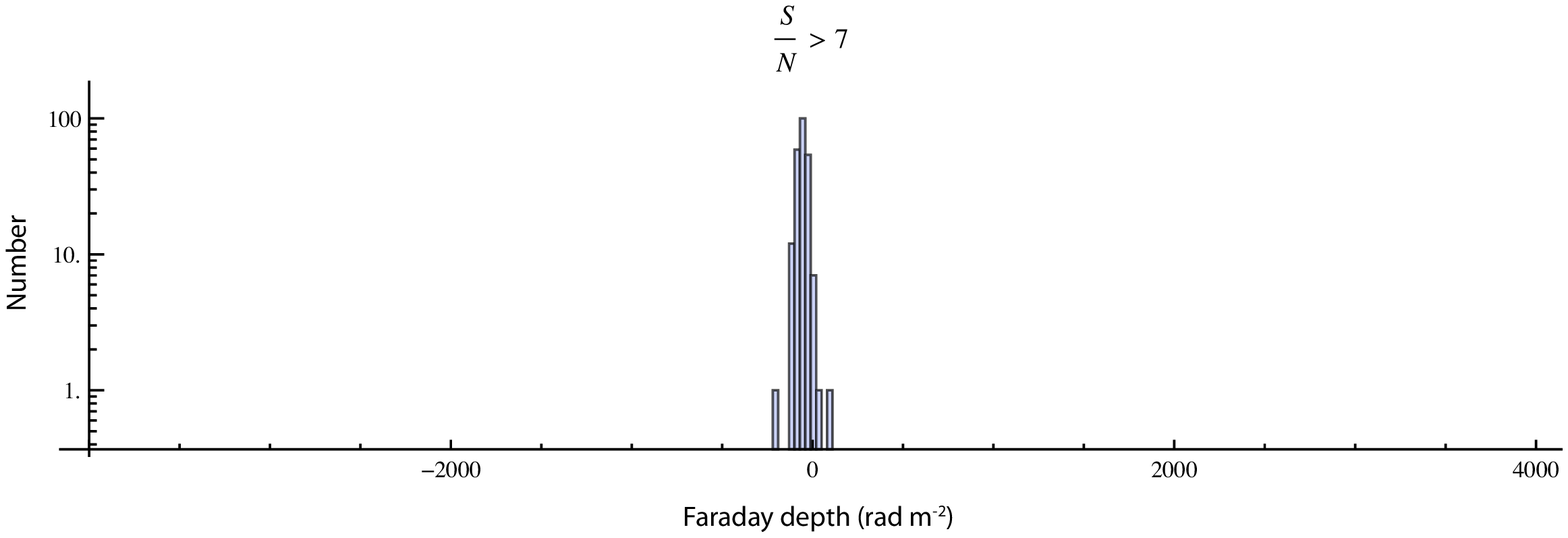}
\end{center}
\caption{Histograms showing the distribution of Faraday depths as a function of the signal-to-noise ratio of the polarization detection.  The logarithmic scaling on the y-axis accentuates bins populated by only a single source.} \label{fig:SNhistogram}
\end{figure}

Figures \ref{fig:SNscat} \& \ref{fig:SNhistogram} show the distribution of derived RMs for varying S/N thresholds.  For the purposes of this analysis we have used only the single most significant Faraday depth derived by RM synthesis; the small number of sources in which there are multiple significant RMs has a negligible influence on the statistics (see \S\ref{sec:complex}).  The signal was derived from the value of $|F|$ at the most significant Faraday depth.  Some RFI had to be flagged in the time based data resulting a small variations in S/N between channels.  This also varied from source to source.  The noise, $\sigma$, is the taken to be the average rms noise per channel for each source divided by the square-root of the weighted number of channels used.   To simplify the theoretical analysis in the following sections we have assumed equal noise per channel, but it was then essential to correct for the effect of this non-uniform noise in the data by including the non-uniform weighting in the S/N estimates and in the estimates of the number of independent channels. We estimate this average rms noise per channel from the observations using the fact that the rms errors add in quadrature.  Feain et al.\,(2009) provides specifics on the estimation of errors for the 8MHz $Q$ and $U$ channels which were interpolated from the correlator output.
The effect of RFI flagging which resulted in non-uniform noise varying from source to source had to be included in the analysis. Even though these effects are small and subtle it is essential to account for them when analysing the polarization statistics.

It is evident that all the observed RMs for high S/N detections are in the range $-150$ to $+50$\,rad\,m$^{-2}$, and we assume that the lower S/N detections are drawn from the same distribution.  Polarization detections with RMs outside this range are most likely to be due to noise.   At the lower S/N regime of interest here, the expected distribution is broadened by noise so we increase the false RM range by $2 \sigma$ and use the range $-210$ to $+110$\,rad\,m$^{-2}$ which is appropriate down to signal strengths of $5 \sigma$.  If Faraday depths associated with false detections are evenly distributed over the depth search range -- as appears to be the case based on Figures \ref{fig:SNscat} \& \ref{fig:SNhistogram} --  there is only a 4\% chance that any given Faraday depth found by RM synthesis will fall in the range of expected real detections by accident. 
Table \ref{tab:detections} gives the distribution of ``true'' and ``false'' polarization detections as a function of S/N.  

Some RMs outside this range could be real (see \S\ref{subsec:empirical}) and while this does not significantly affect the statistical analysis, individual sources with high intrinsic RM are of interest (see \S\ref{sec:complex}).   This population is rare in the flux density range of our sample since only {\it one} of the 309 sources detected above $6 \sigma$ possesses an intrinsic RM magnitude larger than 200\,rad\,m$^{-2}$.

\begin{table*}[ht]
\begin{tabular}{ lccccccccc}
\hline
S/N           &$> 7\sigma$&$> 6\sigma$& $> 5\sigma$ & $> 4\sigma$& $> 3\sigma$&$> 2\sigma$  &  all\\
Approx P(mJy)&  $>0.63$ & $> 0.54$& $> 0.45$ & $> 0.36$ &  $> 0.27$ & $ > 0.18$& \null &\\
\hline
Total number &    	        239 &288  &352  & 488  & 775  & 1003 & 1005 \\
Number with good RM & 239& 287 &346  & 423  & 499  & 527 & 528  \\
Number with false RM &   0   &   1   &  6   &  65    & 276  & 476 & 477 \\
Differential number & \null & 49 & 64 & 136 & 287 & 228 & \null \\
Differential number with & \null  & 48  &  59 & 77 & 76 & 28  & \null \\
 good RM & \null & \null  &  \null & \null & \null & \null  & \null \\
Differential number with & 0  & 1  &  5 & 59 & 211 & 200  & \null \\
 false RM & \null & \null  &  \null & \null & \null & \null  & \null \\
\end{tabular}
\caption{The distribution of polarization detections.  
} \label{tab:detections}
\end{table*}

The one high RM source in the 6$<$S/N$<7$ bin, 132446$-$460218, merits comment.  This source has a clear double peak in the RM synthesis spectrum.  The two peaks have nearly equal polarization: one is at $297\,$rad\,m$^{-2}$ but the second at $-59\,$rad\,m$^{-2}$ does fall within the expected RM range.  The S/N difference in amplitude is $<10$\%.  As discussed at the end of \S5, the statistical treatment of a second RM component in the presence of a significant polarized signal is well beyond the analysis in this paper and this may be a correct but rare detection.

\subsection {Comparison with traditional RM methods}
\begin{figure}[ht]
\begin{center}
\includegraphics[angle=0,width=0.9\textwidth]{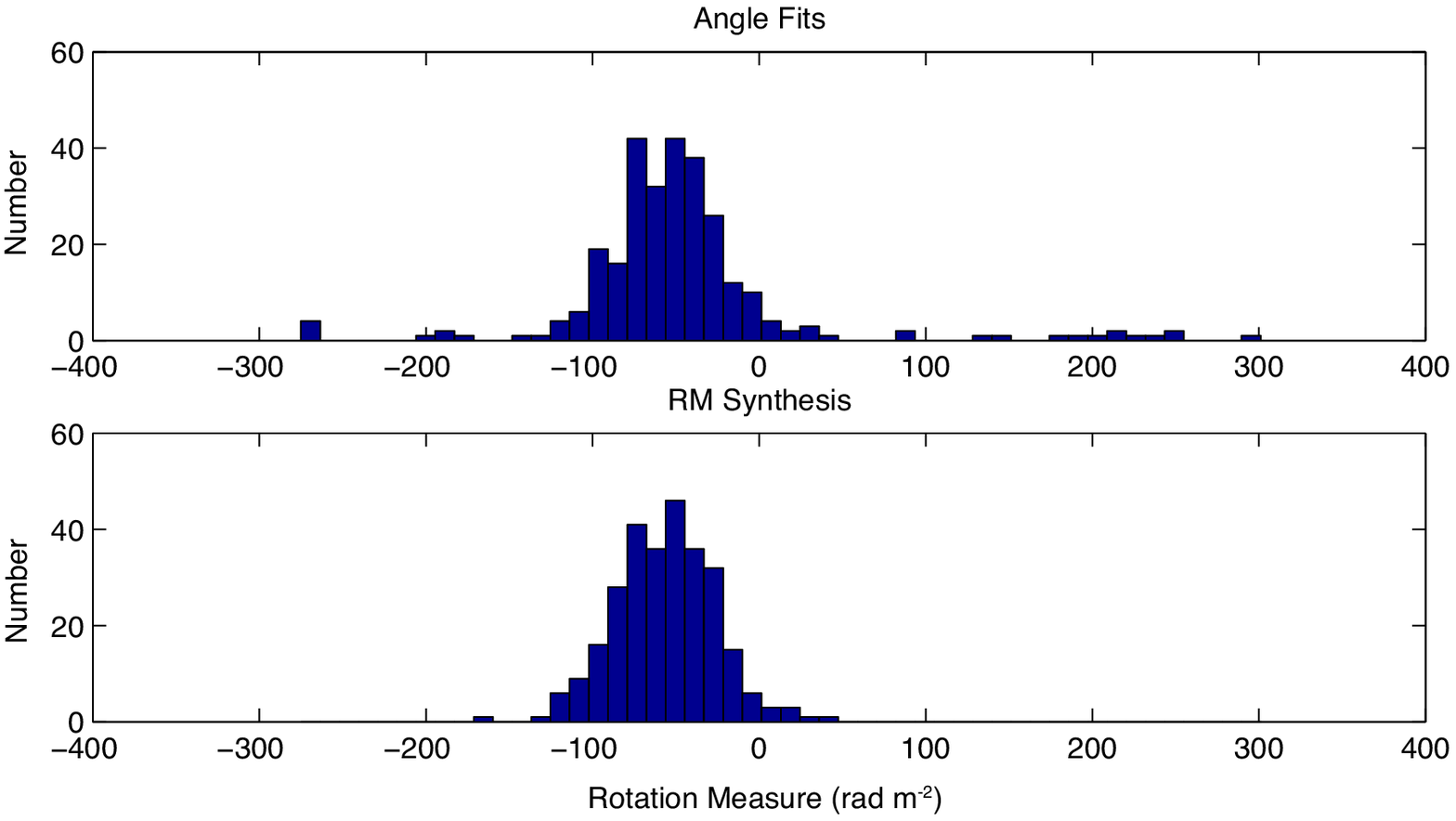}
\end{center}
\caption{Top panel: The distribution of rotation measures derived on the basis of least-squares fits to the polarization position angle measured in each channel.  Bottom panel: the RM distribution of the same sources as determined by RM synthesis using the channelized data.}\label{fig:RMcompare}
\end{figure}

The traditional method to determine the RM has been to determine the position angle of polarization in each frequency channel and then to make a scalar least squares fit to the position angles.  Figure \ref{fig:RMcompare} (top) shows the result of this analysis when applied to sources in the Cen A polarization catalogue.  When compared to the RM synthesis solution for the same sources (bottom panel in Figure \ref{fig:RMcompare}) it can be seen that a small number of spurious high RM values are obtained.  Only those sources with integrated polarization signals S/N$>$7, as identified by RM synthesis, were used in the analysis.  It is evident that there is a tail in the distribution of the Faraday depths derived by fitting polarization position angles.  Of the 251 sources with S/N$>$7 used in the analysis, 16 of these fall outside the range $[-210, 110]$\,rad\,m$^{-2}$.
This can be explained by noting that with the lower S/N in each channel the least squares fit for weak sources can find solutions which are randomly distributed over the RM space included in the least square solution.  This high RM tail is reminiscent of the tail of high RMs sometimes seen in existing RM catalogues e.g. see the comparison of the NVSS catalogs and the Kronberg \& Newton-McGee compilation in Pshirkov et al.\,(2011).   
The small number of genuine high RMs seen in our dataset ($<1$\%) indicates that this class of radio source is rare in surveys at 1.4\,GHz, presumably because sources with high intrinsic RM are more likely to be strongly depolarized.

Taylor et al. (2009) report the rotation measures of 37,543 sources from the NVSS based on VLA snapshot images in two 42\,MHz wide bands centred on 1364.9\,MHz and 1435.1\,MHz.  It is opportune to investigate the accuracy of this catalog given that it currently represents one of the largest resources of rotation measures, and is rapidly being adopted as a de facto standard in analyses of the Galactic and extragalactic magnetic fields (Harvey-Smith, Madsen \& Gaensler 2011; Law et al.\,2011; Opperman et al.\,2011;
Govoni et al.\,2010; McClure-Griffiths et al.\,2010; Schnitzeler 2010; Stasyszyn et al.\,2010).    

We used  the Cen A polarization catalog to investigate the accuracy of this technique used to derive NVSS rotation measures.  The data were obtained at a comparable frequency, and we have collapsed them into two spectral channels with similar centre frequencies of 1364 and 1436\,MHz. As for the NVSS sample, the range of Faraday depths in our sample is sufficiently small that phase wrap ambiguities will not occur\footnote{In the NVSS, the actual rotation measure of a source is related to the deduced value assuming no phase wraps, ${\rm RM}_0$, by the relation, ${\rm RM} = {\rm RM}_0 + (681\, {\rm rad\,m}^{-2}) \, m$, 
where $m$ represents the number of integral phase wraps that occurred between the two observing frequencies.  Taylor et al.\,(2009) assume $|m| \leq 1$, given that RMs of the magnitude required to effect a phase wrap are rare when the source is located off the Galactic plane and away from the inner Galaxy.}.  The RM results for the two-band measurements are in full agreement with the RM synthesis results.  The agreement with the two-band RM synthesis results is unsurprising given that the bands are contiguous.

\subsection{Comparison of RM synthesis and least squares fits} \label{subsec:RMsynthVsBruteForce}
Although RM synthesis outperforms traditional frequency-channel-based position-angle fitting algorithms such as the ones discussed above, it remains to be seen whether it makes the best use of the polarization information available.  We address this by considering the output of a brute force minimum least squares search for detected polarization at any RM on all the sources, subject to the assumption that the source contains only a single polarized component that is assumed to have a flat spectrum.  To be specific, we implemented a code that seeks the parameters $P$, $\chi_0$ and $\phi$ (polarization amplitude, position angle and Faraday depth) that minimize the penalty function,
\begin{eqnarray}
\chi^2 = \frac{1}{2 N - 3} \sum_j^N \frac{[Q_j - {\rm Re}[p_j(P,\chi_0,\phi) ]^2}{\sigma_Q^2} + \frac{[U_j - {\rm Im}[p_j(P,\chi_0,\phi) ]^2}{\sigma_U^2},
\end{eqnarray}
where the model for the polarization of the $j$th channel is, 
\begin{eqnarray}
p_j(P,\chi_0,\phi) = P \exp[ 2 i (\chi_0 + \lambda_j^2 \phi)].
\end{eqnarray}

\begin{figure}[ht]
\begin{center}
\begin{tabular}{cc}
\includegraphics[angle=0,width=0.45\textwidth]{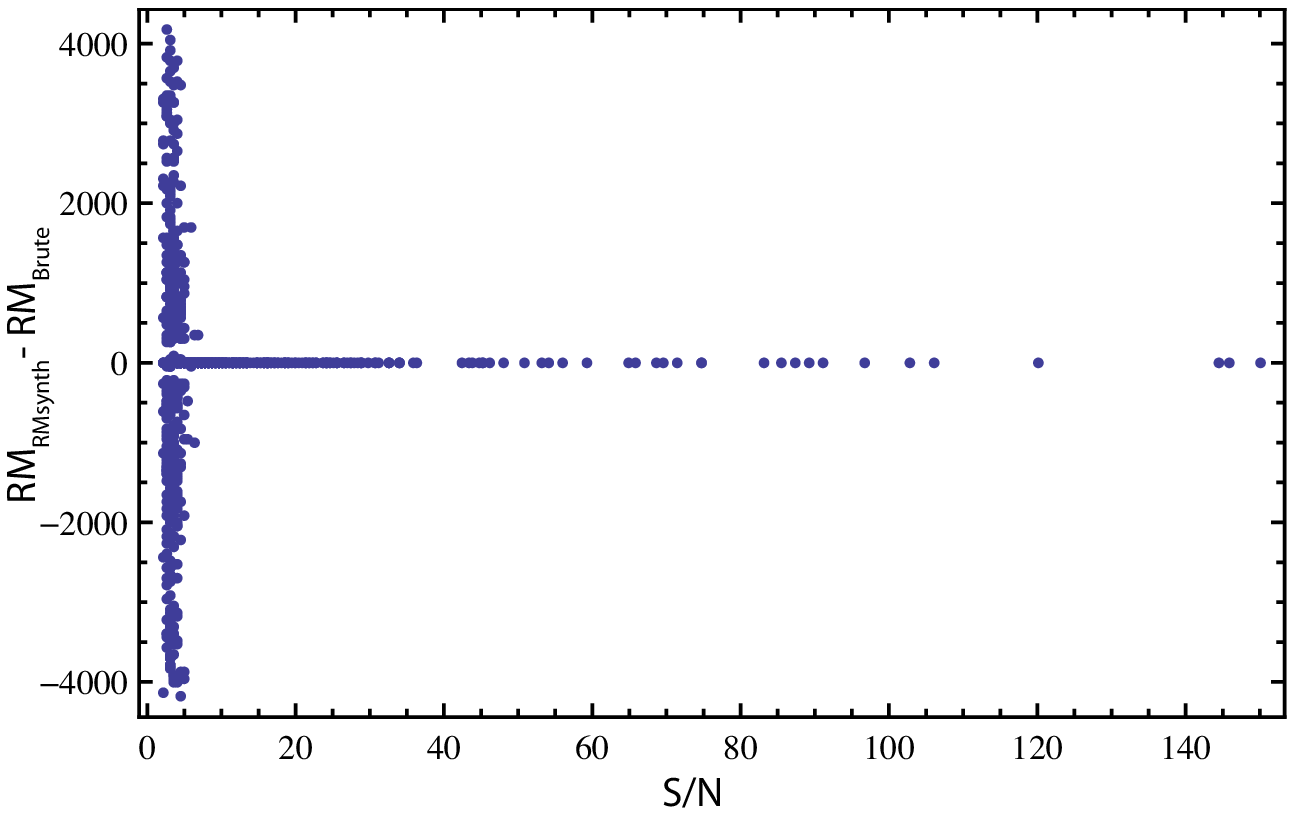} &
\includegraphics[angle=0,width=0.45\textwidth]{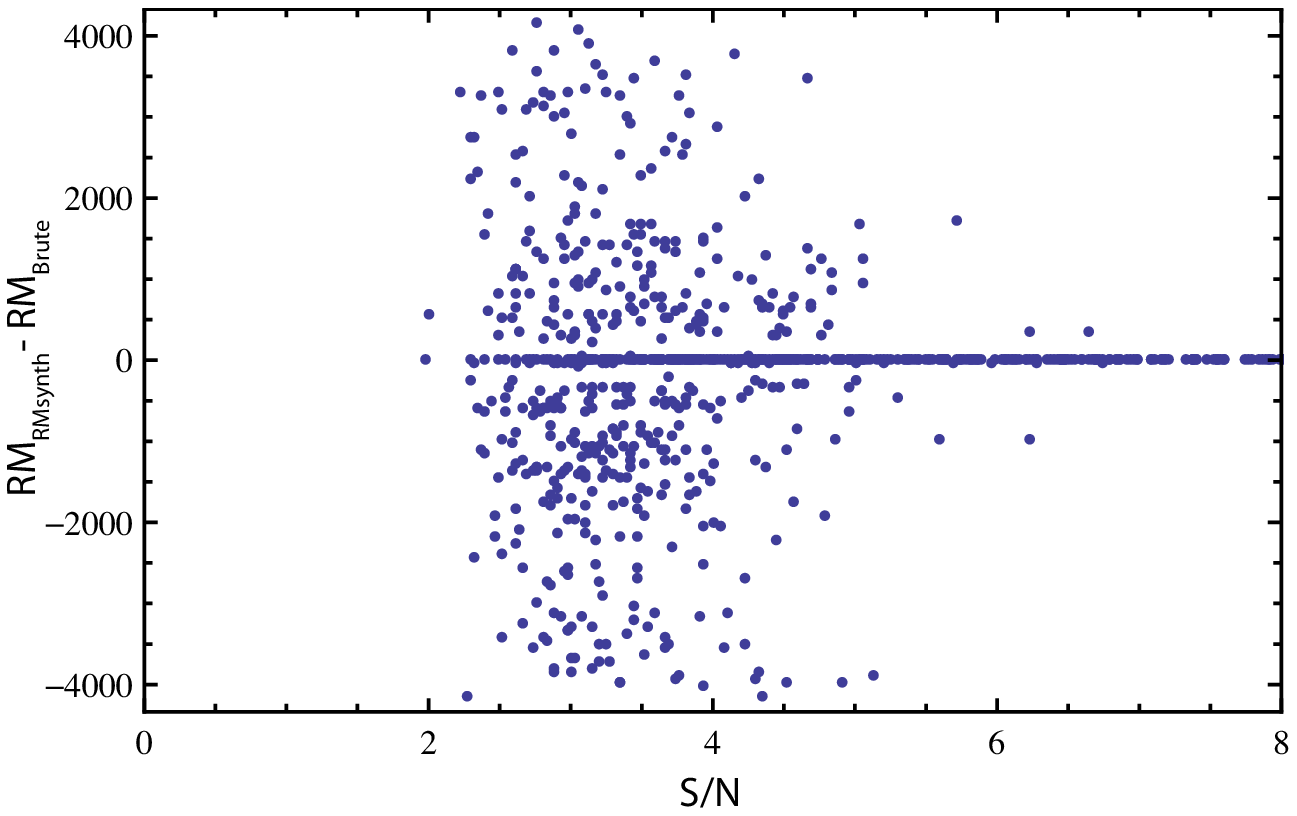} \\
\includegraphics[angle=0,width=0.45\textwidth]{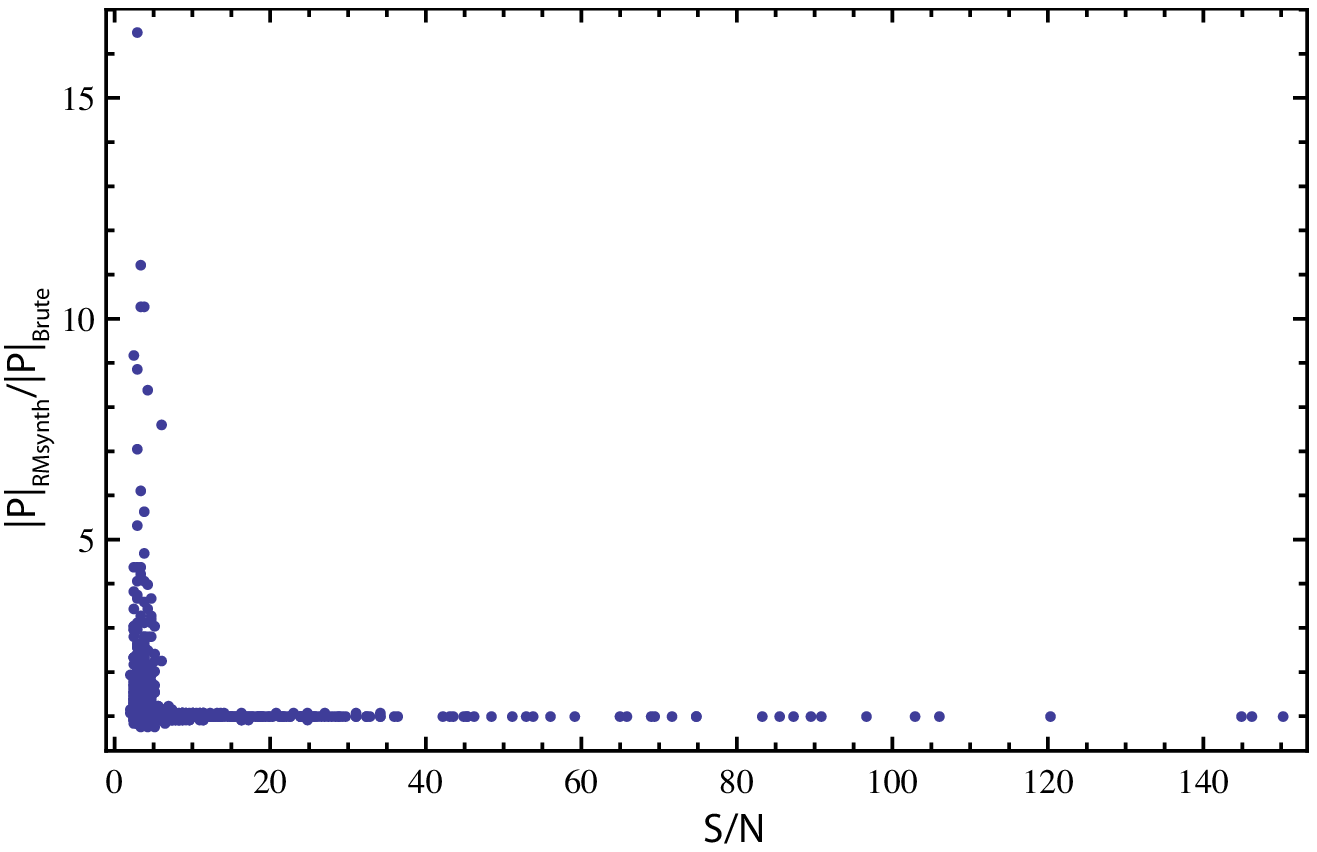} & 
\includegraphics[angle=0,width=0.45\textwidth]{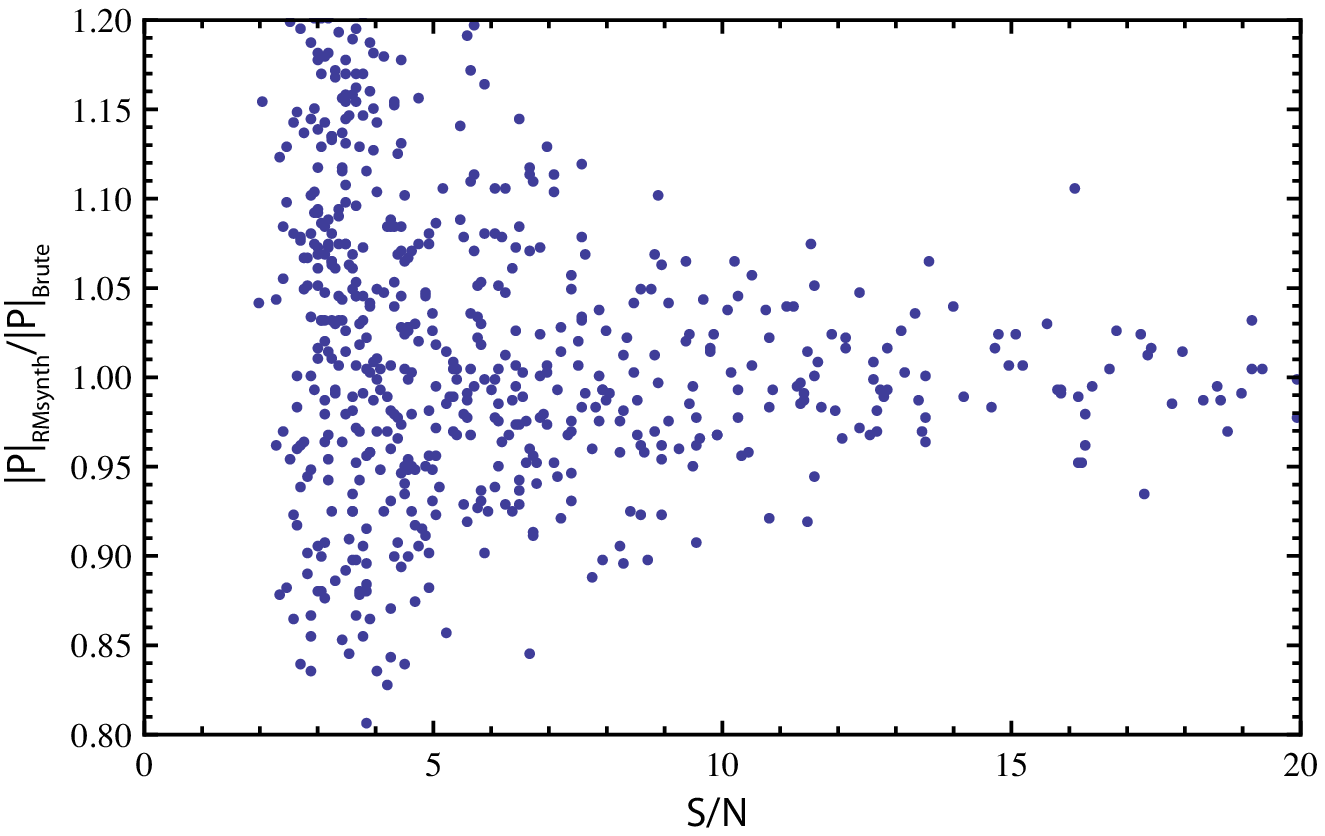} \\
\end{tabular}
\end{center}
\caption{Comparison of the RM and polarization amplitude derived by RM synthesis compared to a brute force least squares fit to the polarization data, as described in \S\ref{subsec:RMsynthVsBruteForce}.}\label{fig:brute force}
\end{figure}

Figure \ref{fig:brute force} compares the solutions from the RM synthesis with the brute force least squares fit approach.  The two methods agree very well in both RM and polarization amplitude.  At high S/N the solutions are identical.  The increased scatter at lower S/N is not surprising as the two methods find different solutions, but the gap (along the Faraday depth axis) surrounding the identical solution indicates that both methods converge to the identical solution if the solutions are close.  There is no systematic change in the ratio of signal amplitude between the two methods with S/N, and the average ratio for the two methods is 1.0.  

The behaviour of the solutions for the sources with a ``false'' detection based on the RM has a number of undesirable properties.  The brute force solution has a much narrower distribution of RMs at low S/N ($\pm$ 500 rad m$^{-2}$). The amplitudes of the bad solutions are lower and the reduced $\chi^{2}$  worse indicating that it does not find the global minimum for noise-dominated signals.

In \S\ref{subsec:sigPlusnoise} it is demonstrated that the noise does fall as $N^{-1/2}$ for RM synthesis.  The na\"ive expectation would be that it falls similarly for traditional methods, since the estimated standard deviation in the slope of a regression line scales as $N^{-1/2}$.  However, in practice this dependence is more complicated for traditional methods in a way that is hard to analyse; $n \pi$ phase ambiguities in the position angle introduce additional degrees of freedom into the fitting problem, and the fit error depends in detail on how the channels are spaced.

\section{The statistics of polarized sources} \label{sec:re-analysisAndNoise}

Determination of the distribution of polarization amplitudes intrinsic to a set of sources involves disentangling the contribution of noise from the polarization amplitudes deduced by RM synthesis.  
This is most important for sources whose intrinsic polarization is near the noise level.  

We are motivated to examine the noise statistics in detail with a view to accounting for its effects in the distribution of polarization amplitudes measured in a set of sources.  Specifically, the objective is to evaluate the quantity $p( P {\vert} P_0)$, namely the probability of obtaining an observed polarization amplitude $P$ conditioned on the amplitude of the true intrinsic polarization, $P_0$.  This distribution function is the basis for deriving the probability distribution of intrinsic (i.e.\,``noise-corrected'') polarization amplitudes based on a set of measured polarization amplitudes determined by RM synthesis.  We conclude this section by applying the formalism to investigate the distribution of polarization amplitudes in the Cen A dataset at low polarization amplitude levels.

\subsection{The polarization distribution of noise} \label{subsec:noise}

Here we derive the distribution of polarization amplitudes recovered by RM synthesis from pure noise.
The input measurements $Q_j$ and $U_j$ are therefore assumed to possess the properties of thermal noise, and are independent and normally distributed with zero mean and noise per spectral channel $\sigma$.

Recall that RM synthesis derotates the channelized measurements of $Q$ and $U$ for a range of different trial rotation measures, $\phi$.  The resultant summed (complex) number is 
\begin{eqnarray}
{\cal P}  &=& \frac{1}{N} \sum_j^N P_j e^{i \psi_j} e^{2 i \lambda_j^2 \phi}. \label{RMsum}
\end{eqnarray}
The polarization vector derived by RM synthesis is the value of ${\cal P}$ for which the choice of $\phi$ gives a maximum in the value of $P$.  The task here is to determine the distribution of the amplitude of ${\cal P}$.  If $\phi$ is given it is straightforward to recover the polarization distribution.  However, an additional complication arises because the RM synthesis algorithm is allowed the extra degree of freedom to chose any value of $\phi$ which maximises $P$.  The freedom to rotate the measured polarization vectors over a range of trial values of $\phi$ ensures that the polarization vector amplitude returned by RM synthesis is at least as large as that found in the $\phi=0$ case.  This is because the algorithm may derotate the measured polarization vectors with any value of $\phi$ which improves the coherence of the sum in eq.\,(\ref{RMsum}).

There is a direct analogy between the noise characteristics of RM synthesis and those encountered in interferometric radio imaging, which involves deriving the distribution of noise amplitudes from (complex) visibilities (see Section 9.3 of Thomson, Moran \& Svenson 2001).   In both cases one is attempting to ``wind-up'' complex valued quantities whose real and imaginary components are both normally distributed.  In the case of RM synthesis, they are wound up for various values of Faraday depth.  In the interferometric imaging case the complex-valued interferometric visibilites are searched across a range of fringe rotation values, corresponding to the possible position of the source.  Thus the two treatments are mathematically identical.  We elaborate on this connection in the analysis below.

\subsubsection{Constant $\phi$} \label{subsubsec:ConstPhi}
First consider the simple, well-known case in which we explicitly hold the rotation measure at a fixed value.  Since the input data is purely noise-like the actual value of $\phi$ is unimportant: any noise-like polarization vector which is transformed by derotation to any nonzero value of $\phi$ still possesses the same statistical noise-like properties.  Here, for the purpose of simplicity, we set $\phi=0$.  The polarization found by RM synthesis is,
\begin{eqnarray}
{\cal P}_{\phi=0} = \frac{1}{N} \left[ \sum_j^N Q_j + i \sum_j^N U_j \right].
\end{eqnarray}
The statistics of the sum of $N$ independent, normally distributed random variates is also normally distributed, with standard deviation $\sigma_{Q,U} \sqrt{N}$.  Thus $S_Q=N^{-1} \sum_j^N Q_j$ also follows a normal distribution with zero mean and standard deviation $\sigma_{Q,U}/ \sqrt{N}$.
An identical relation holds for $S_U=N^{-1} \sum_j^N U_j$.  It is then straightforward to show that 
the probability distribution for the polarization amplitude when there is no signal present and $\phi=0$ is just a Rayleigh distribution,
\begin{eqnarray}
p_P(P)dP =  \frac{P}{\Sigma^2} \exp \left( - \frac{P^2}{2 \Sigma^2} \right) d P, \label{PRayleigh}
\end{eqnarray}
where we define $\Sigma^2 = \sigma^2/N$.  This result is well known in the analogous interferometric imaging case pertaining to noise in VLBI observations, where it corresponds to the statistics of the fringe amplitude for a specific fringe rotation when no signal is present (Thomson, Moran \& Swenson 2001, p318).

\subsubsection{Unconstrained $\phi$} 
Now consider the general case in which RM synthesis performs a fit to the polarization data that is unconstrained in Faraday depth, $\phi$.  This consists of two separate steps.  The first is to find the specific value of $\phi$ which maximizes the value of $P$ for a given dataset.  The second step is to find the distribution of $P$ corresponding to these maxima.

The square of the polarization amplitude is
\begin{eqnarray}
{\cal PP^*} &=& \sum_{k,m}^N P_k P_m e^{i (\psi_k - \psi_m)} e^{2 i \phi (\lambda_k^2 - \lambda_m^2)} \nonumber \\
&=&  \sum_{k,m}^N P_k P_m \cos [ \psi_k - \psi_m + 2 \phi (\lambda_k^2 - \lambda_m^2)].
\end{eqnarray}
The second line follows because the sine function has odd symmetry, so the imaginary part of every pair $k,m$ in the sum is cancelled by every pair $m,k$.  

The rotation measure $\phi$ should obey the following condition at the maximal value of $P$, 
\begin{eqnarray}
\frac{\partial ({\cal PP^*})}{\partial \phi} = 0 = \sum_{k,m}^N 2 i (\lambda_k^2 - \lambda_m^2) P_k P_m e^{i (\psi_k - \psi_m)} e^{2 i \phi (\lambda_k^2 - \lambda_m^2)}.
\end{eqnarray}
Since the imaginary terms are antisymmetric with respect to exchange of the indices $k$ and $m$, we see that imaginary part of this sum is zero for any value of $\phi$.  So only the real part of the sum 
yields a nontrivial constraint on $\phi$:
\begin{eqnarray}
0 = \sum_{k,m}^N (\lambda_k^2 -\lambda_m^2) P_k P_m \sin [\psi_k - \psi_m + 2 \phi (\lambda_k^2 - \lambda_m^2) ].
\end{eqnarray}
The foregoing relations show that the distribution of ${\cal P}$ exhibits a detailed dependence on the spectral coverage (i.e.\,the values of $\lambda_k^2$).  This indicates that the exact location and amplitude of the noise peak depends on the $\lambda_k^2$, and that the noise distribution for any given spectral setup is best found empirically.

It is possible to construct a simple approximate analytic formulation independent of the detailed spectral coverage.  This result has already been derived in the analogous case of the noise properties in VLBI images, where one is concerned with the statistics of the fringe amplitude across  (Thomson, Moran \& Swenson 2001, pp319-211).  We briefly summarize the argument here and place it in the context of RM synthesis. 

The statistics of the polarization amplitude at any one value of $\phi$ obey a Rayleigh distribution.  At any given value of $\phi$ the probability that we detect a polarization amplitude lower than some prescribed threshold amplitude, say $Z$, is the cumulative probability distribution corresponding to eq.\,(\ref{PRayleigh}), namely, $p(Z  \leq  P) = 1 - \exp \left( - {Z}^2/2 \Sigma^2 \right)$. If the RM synthesis search is conducted over a large range of possible Faraday depths, $-\phi_0 < \phi < \phi_0$, and the width of the amplitude of the RMTF is approximately $\Delta \phi$, there are $N \approx 2 \phi_0/\Delta \phi$ independent Faraday depth ``pixels'', or trials, over which the RM search is being performed.  The probability that we do not detect a polarization amplitude above some amplitude $Z$ after these $N$ trials is $p^N$.  The probability of finding $Z \leq |{\cal P|}$ is just the culmulative probability distribution function of the amplitude of the polarization vector,
\begin{eqnarray}
{\rm CDF}_{n} (Z ) = \left[ 1 - \exp \left( - \frac{Z^2}{2 \Sigma^2} \right) \right]^N,
\end{eqnarray}
where the subscript $n$ emphasises that the distribution is associated with a purely noise-like signal.
The corresponding probability density function is,
\begin{eqnarray}
p_n(Z)  = \frac{N Z}{\Sigma^2} \left[ 1 - \exp \left( - \frac{Z^2}{2 \Sigma^2} \right) \right]^{N-1} \exp \left( - \frac{Z^2}{2 \Sigma^2} \right) . \label{pNoise}
\end{eqnarray}
This expression describes the expected distribution of the polarization amplitude when the input data are purely noise-like and in which the rotation measure is searched over approximately $N$ independent Faraday depth pixels.  It is identical the characteristics of the noise in a search for fringes from $N$ independent samples (Thomson, Moran \& Swenson 2001, eq 9.57).  
The exact noise distribution pertaining to the Cen A dataset is discussed at the conclusion of this section.  We pre-empt this by noting that we find empirically that the noise distribution for the Cen A observations (shown in Fig.\,\ref{fig:simsPamp}) differs slightly from the analytic formula derived here.

The behaviour of the noise distribution has a number of important implications for the statistics of polarization amplitudes determined by RM synthesis.  Equation (\ref{pNoise}) shows that the distribution of polarization amplitudes derived from noise narrows as $N$ increases (for constant $\Sigma$), and the peak of the distribution shifts to increasing values of $P/\Sigma$.  The location of the peak is determined by the solution to the transcendental equation,
\begin{eqnarray}
P^2 \left[ N -\exp(P^2/2 \Sigma^2) \right] = \Sigma^2 \left[ 1 - \exp( P^2/2 \Sigma^2) \right].
\end{eqnarray}
This has solutions $P/\Sigma = 1.93, 2.24, 2.84, 3.07$ and $3.54$ respectively for $N=5,10,50,100$ and $500$.   

Thus the threshold signal-to-noise, $P/\Sigma$, for the believability of a polarization signal extracted using RM synthesis exhibits an important, albeit weak, dependence on the number of independent trials, $N$.   The distribution peaks sharply on the low $P/\Sigma$ side of the distribution, implying that one expects very few polarization amplitudes with signal-to-noise values less than $2.0$ once $N$ exceeds $10$.  For the analysis in this paper $N=30$.

\subsection{The polarization distribution of a nonzero polarization signal containing noise} \label{subsec:sigPlusnoise}

There is an important distinction between the polarization distribution obtained from a purely noise-like signal from RM synthesis and the polarization distribution derived when an intrinsic (nonzero) polarization signal is present.  The distinction arises because, in the absence of an intrinsic polarization signal, the RM synthesis algorithm picks out the polarization vector associated with the Faraday depth that maximises the length of the ``Faraday-derotated'' noise vector.   However, this freedom to optimise the length of the polarization noise vector is removed provided the intrinsic signal is sufficiently large that the RM synthesis code finds the highest amplitude at the correct Faraday depth of the intrinsic signal (i.e.\,there is no other Faraday depth at which a purely noise-like signal yields a higher polarization amplitude).  In this instance the RM synthesis code is constrained to derotate the signal at the Faraday depth of the intrinsic polarization signal.

We compute the probability distribution of the polarization signal subject to the assumption that the intrinsic polarization is nonzero and that the RM synthesis code detrotates the signal at the Faraday depth associated with the intrinsic signal\footnote{The contribution of noise, at the Faraday depth of the intrinsic signal may, in principle, slightly alter the depth at which the peak amplitude occurs, and hence the Faraday depth reported by the RM synthesis algorithm.  This may, in turn, influence the statistics of $P$.  However, this is a small effect in most practical situtations.  
If the Faraday depth derived by RM synthesis is sufficiently close to the correct depth it will not alter the statistics of $P$; this is the case provided that the S/N is sufficient that the error in Faraday depth is a small fraction of the width of the rotation measure transfer function, $R(\phi)$ (i.e. the S/N $\gtrsim 2$).  Winding up an intrinsic polarization signal $P_0$ at a depth that is wrong by an amount $\Delta \phi$ yields an intrinsic signal of amplitude $\sim P_0 \cos  \lambda_0^2 \Delta \phi $, so the error in the amplitude of the intrinsic polarization vector,  $\approx P_0 ( \lambda_0^2 \Delta \phi)^2/2$, is quadratic in $\lambda_0^2  \Delta \phi$.  This is a small quantity since, for S/N $\gtrsim 2$ the peak is localised to within the width of $R(\phi)$, one has $\lambda_0^2  \Delta \phi \ll 1$, so that $p( Z | P_0 ) \approx  p( Z | P_0  \cos  \lambda_0^2 \phi)$. We see in the following subsection that the condition S/N $\gtrsim 2$ is always satisfied.}.   
Suppose we have a nonzero intrinsic signal whose Faraday de-rotated channel measurements of the linear polarization are ${\cal P}_0=(Q_0, U_0)$.  Associated with each spectral channel there is also a noise contribution.  The noise follows a Gaussian distribution.  It is also uncorrelated with the intrinsic signal and thus its statistical properties remain identical after de-rotation at the Faraday depth of the intrinsic signal.  We write the derotated noise polarization contribution for the $i$th spectral channel as $\Delta Q_i$ and $\Delta U_i$.  Then the polarization vector derived by RM synthesis is
\begin{eqnarray}
{\cal P} = \left(  Q_0 + S_Q  ,  U_0 + S_U \right).
\end{eqnarray}
where $S_Q = N^{-1} \sum_i^N \Delta Q_i$ and $S_U = N^{-1} \sum_i^N \Delta U_i$ (as defined previously).  
The polarization vector, ${\cal P}$, therefore obeys the statistics of a random walk with a constant offset ${\cal P}_0=(Q_0,U_0)$.  Thus the polarization amplitude in the presence of an intrinsic signal follows the Rician distribution,
\begin{eqnarray}
p_{s}(Z=P \vert P_0 ) = \frac{Z}{\Sigma^2} \exp \left[  - \frac{Z^2 + P_0^2}{2 \Sigma^2}\right] 
I_0 \left( \frac{Z | P_0|}{\Sigma^2} \right), \label{PdistnSignal}
\end{eqnarray}
where $I_0(x)$ is the modified Bessel function of the first kind of order zero and the subscript $s$ denotes that the distribution is valid only when a nonzero polarization signal is present so that the signal is derotated at the Faraday depth corresponding to the intrinsic signal, ${\cal P}_0$.  
The associated cumulative probability distribution function (i.e. the probability of $Z \leq  P$) is
\begin{eqnarray}
{\rm CDF}_s (Z  \vert  P_0 ) = 1  - Q_1 \left( \frac{ P_0}{\Sigma}, \frac{Z}{\Sigma} \right), 
\end{eqnarray}
where $Q_m(a,b)$ is Marcum's Q function.
This result is analogous to the probability distribution of the interferometric fringe amplitude in which one measures a nonzero signal in the presence of noise (Thomson, Moran \& Swenson 2001, eq.\,9.37).

Some remarks about the treatment of polarization noise bias are in order.  
The polarization amplitude distribution in eq.\,(\ref{PdistnSignal}) demonstrates that the measured amplitude is, {\it on average}, biased by noise.  In particular one has,
\begin{eqnarray}
\langle P^2 \rangle =  P_0^2 + 2 \Sigma^2.
\end{eqnarray}
Thus the appropriate noise bias to subtract from a single measurement of $P^2$ derived on the basis of RM synthesis is twice the channel-integrated noise variance, $\Sigma^2$.  While correct on average, bias subtraction is clearly inexact for any given source; in as much as the noise can assume a range of values, so is there a range of values of intrinsic $P_0$ that are consistent with a given measured $P$.   Our analysis shows that it is entirely inappropriate to simply remove a noise bias from the data and treat the resulting polarization amplitudes as if they were distributed with a Gaussian error distribution.

\subsection{Likelihood of detection} \label{subsec:detectionlikelihood}
In the two forgegoing subsections we have derived the probability distributions of the polarization amplitude both when a signal is absent, and when one is present and its amplitude is sufficiently large that it dominates the noise (i.e.\,provided $P_0$ exceeds some threshold).  

Here we deduce the probability distribution of the polarization amplitude that is valid for {\it all} $P_0$.  The key to this derivation is that there is a nonzero probability that the value of $P$ drawn from the signal distribution $p_s$ is lower than the value that would be drawn from the noise distribution $p_n$.  Under this circumstance, the value of $P$ picked by the RM synthesis algorithm is the greater of the two values, namely that drawn from the noise distribution.  

The probability of obtaining a polarization amplitude of $Z=P$ is the sum of (i) the probability of drawing $Z$ from the noise distribution and requiring that its value exceeds the amplitude drawn from the signal distribution, and (ii) the probability that $Z$ was instead drawn from the signal distribution and requiring its value exceeds the amplitude drawn from the noise distribution.  Thus, one obtains the combined probability distribution, 
\begin{eqnarray}
p_{\rm c} ( Z = P  \vert \, P_0 ) = p_n (Z)  {\rm CDF}_s( Z | P_0 ) + p_s (Z | P_0)  {\rm CDF}_n(Z).
\end{eqnarray}
This distribution is valid for any value of $P_0$. It obeys the expected property that at low $Z$ and $P_0$ the distribution is dominated by the noise distribution, $p_n(Z)$, because ${\rm CDF}_n(Z)$ is almost zero in this region, while it is dominated by $p_s (Z | P_0)$ at high $Z$ because $p_n(Z)$ is almost zero in this domain.

\begin{figure}[ht]
\begin{center}
\begin{tabular}{c}
\includegraphics[angle=0,width=0.85\textwidth]{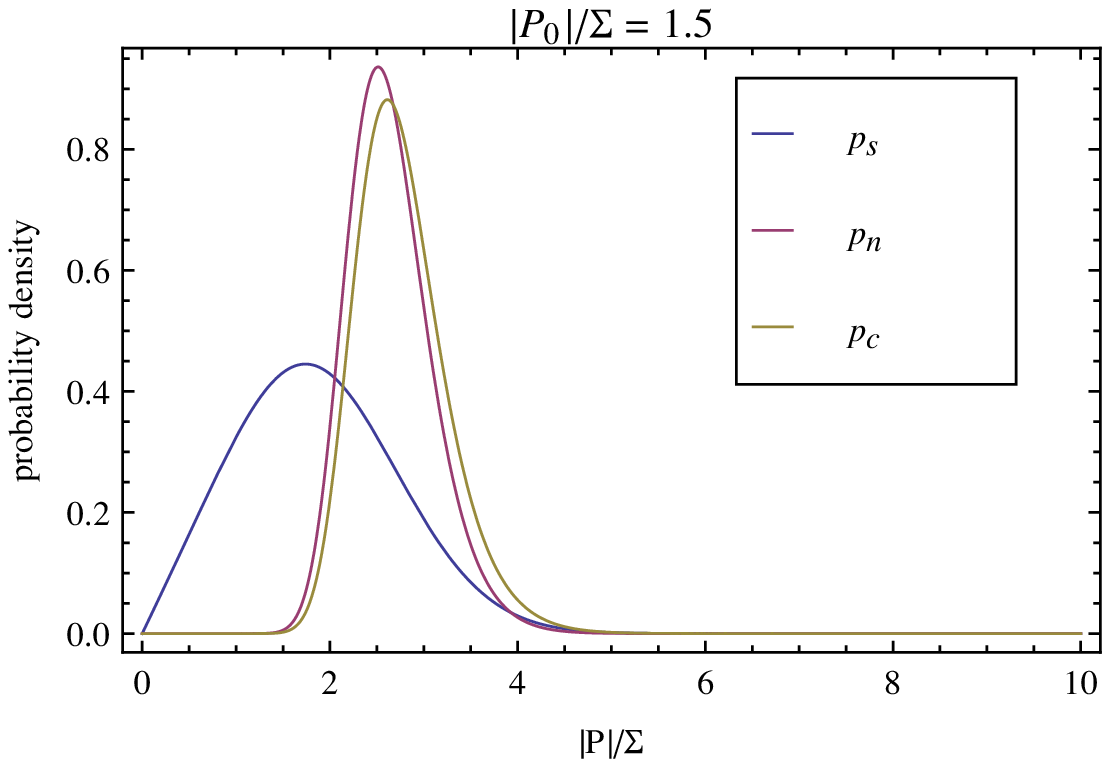} \\
\includegraphics[angle=0,width=0.85\textwidth]{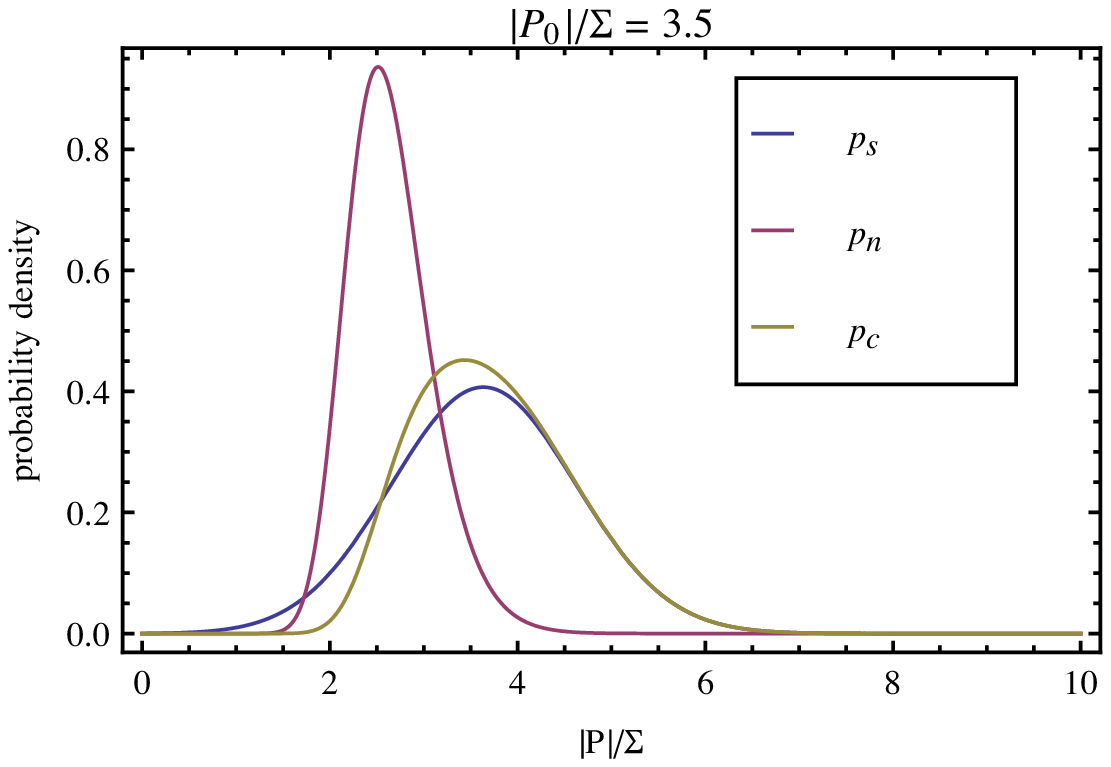} \\
\end{tabular}
\end{center}
\caption{An illustration of the various distributions $p_s$, $p_n$ and $p_c$ as a function of $P$ for $N=20$ and (top) $| {\cal P}_0| = 1.5/\Sigma$ and (bottom) $|{\cal P}_0 |/\Sigma = 3.5$.  For values of $|{\cal P}_0| \lesssim 1 $ the distribution closely resembles the noise distribution, $p_n$, while for $|{\cal P}_0|/\Sigma \gtrsim 1$ it closely resembles the signal probability distribution, $p_s$.} \label{fig:P0alldistn}
\end{figure}

By way of illustration, Fig.\,\ref{fig:P0alldistn} compares the combined probability distribution for some specific choices of $P_0$ against $p_s$ and $p_n$.

One can extend the foregoing logic to determine the likelihood that a signal found by RM synthesis at a given amplitude, $Z$, is a genuine signal.  The probability that the signal has been derotated to the correct value of $\phi$ is,
\begin{eqnarray}
p_{\phi \hbox{ correct}}(Z | P_0) = \frac{p_s (Z | P_0)  {\rm CDF}_n( Z)}{p_c (Z | P_0)}.
\end{eqnarray}
It is apparent that, even for small values of $Z$ and $P_0$, there is a nonzero probability that the RM synthesis code deduces the correct Faraday depth of the signal.  

This is borne out by the Cen A data: close inspection of the lower panel in Fig.\,\ref{fig:SNscat} indicates that even at S/N $\sim 3.5$ the RM synthesis algorithm preferentially finds signals within the range of physically acceptable values of $\phi$.

\subsection{Deduction of unbiased polarization amplitude source counts}
The foregoing formalism provides an obvious prescription to determine the distribution of intrinsic polarization amplitudes, $N(P_0)$ based on the measured distribution of polarization amplitudes as deduced by RM synthesis, namely $N(P)$:
\begin{eqnarray}
N( P) = \int d P_0 \,\, p ( P \vert  P_0 ) N (P_0).
\label{SrcCounts}
\end{eqnarray}
The solution of eq.\,(\ref{SrcCounts}) to find $N(P_0)$ is formally an ill-posed problem, but is it of a form that is often encountered in numerical optimization theory.

It is instructive to demonstrate how the solution of eq.(\ref{SrcCounts}) is readily ammenable to numerical solution.  The distribution of intrinsic polarization amplitudes, $N( P_0)$ must be solved by sampling onto a grid with resolution, say, $\Delta | P|$.  We represent this distribution by the vector $y_j$.   The corresponding distribution of measured polarization amplitudes is discretized onto a grid in $P$ with identical resolution, and this is represented by the vector $x_i$.   If one similarly discretizes $p_c (Z \, | \, P_0)$ into a matrix whose $i,j$th element corresponds to the parameters $Z_i$ and $ {P_0}_j$, with resolution  $\Delta  P $ in both dimensions, the integral in eq.(\ref{SrcCounts}) can be cast in the form,
\begin{eqnarray}
x_i = \sum_j A_{ij} y_j \Delta P. \label{SrcCountsMatrix}
\end{eqnarray}
This class of problems is ammenable to solution by iterative deconvolution and expectation-maximization algorithms.

A simpler but more robust approach is to forward-model the expected distribution $N(|{\cal P}_0|)$ to determine whether it is consistent with the observed amplitude distribution.  

\subsection{Application to Centaurus A data} \label{subsec:empirical}

As an illustration of the foregoing formalism, we apply it to determine the distribution of intrinsic polarization amplitudes in the Cen A dataset.   

In order to model the effect of noise in this dataset precisely, we determined the noise distribution $p_n$ empirically based on the specific spectral coverage of the ATCA observations.  The same RM synthesis code used to process the Cen A data was employed.  We synthesised $50,000$ purely noise-like sources where, for each source, synthetic $Q$ and $U$ data for each spectral channel were drawn from a normal distribution of zero mean and unit standard deviation, and assigned to spectral channels with frequencies identical to those used in the Cen A observations.  

\begin{figure}[ht]
\begin{center}
\begin{tabular}{c}
\includegraphics[angle=0,width=0.78\textwidth]{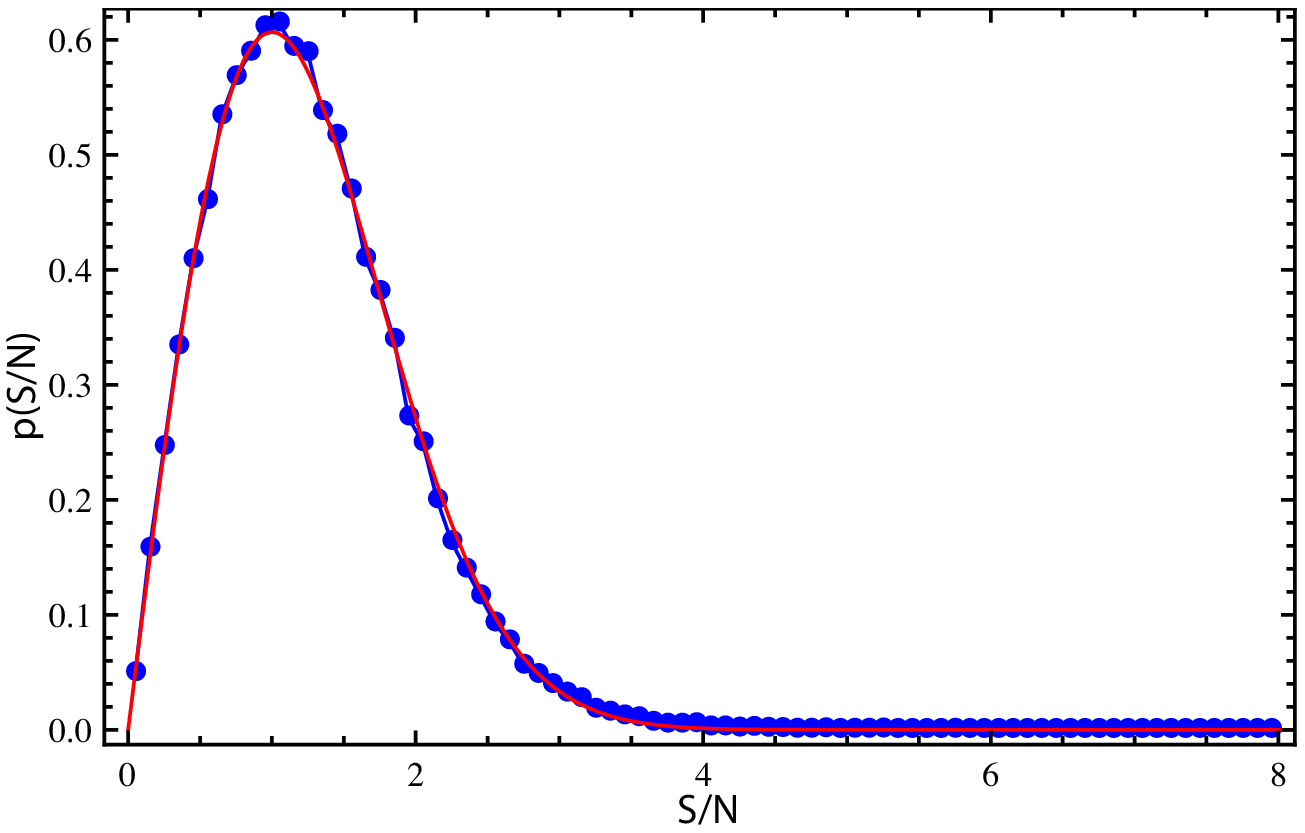} \\
\includegraphics[angle=0,width=0.78\textwidth]{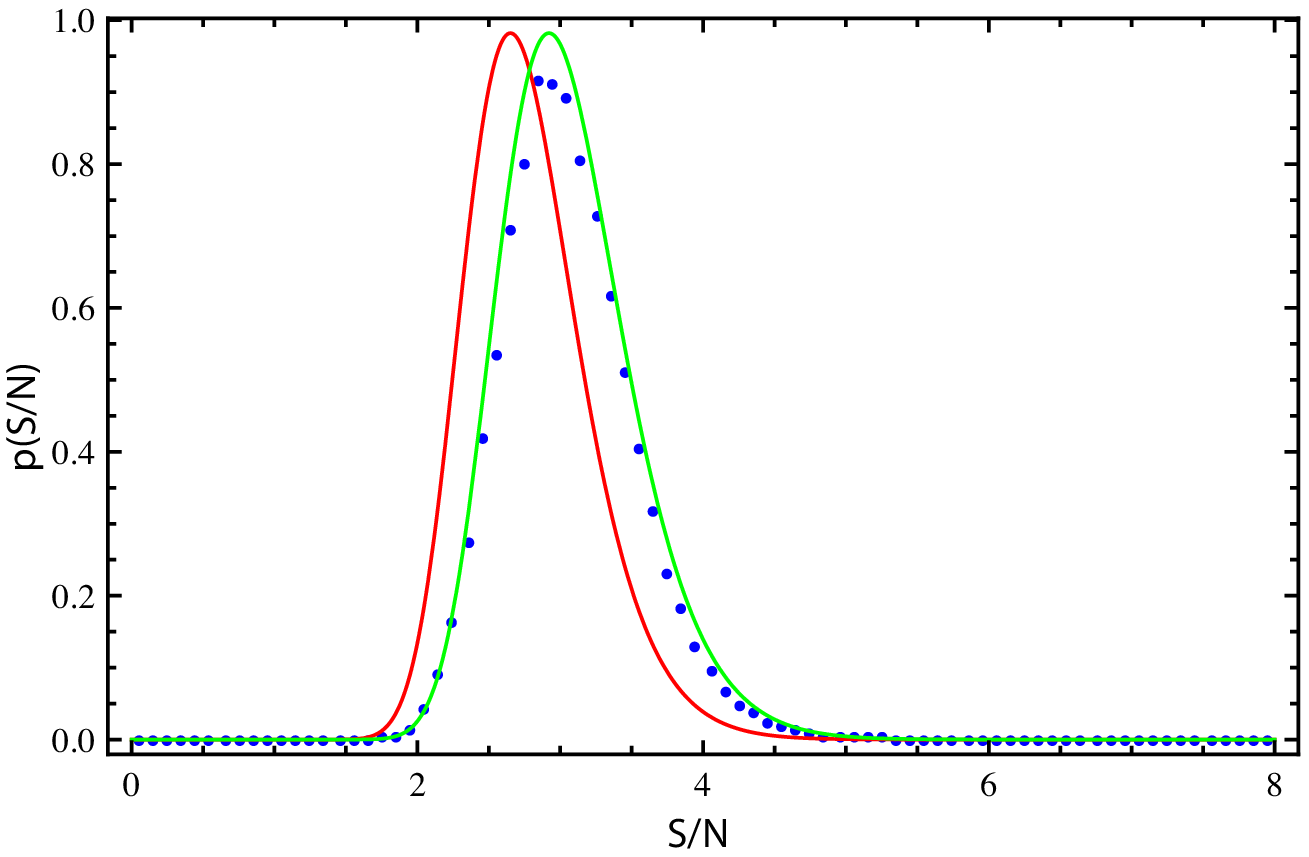} \\
\end{tabular}
\end{center}
\caption{The amplitude of the polarization vector recovered by RM synthesis from a purely noise-like signal for the spectral channel setup used in the Cen A observations.  The upper panel depicts the distribution when the rotation measure is constrained to a fixed value, here $\phi=0$.  The overplotted curve (red) shows the predicted distribution (see \S\ref{subsubsec:ConstPhi}).  The lower panel depicts the distribution of recovered polarization amplitudes when the rotation measure is allowed to fall anywhere in the range allowed for the polarization extracted from the real data, namely $-4000\,\hbox{rad\,m}^{-2} \leq \phi \leq 4000$\,rad\,m$^{-2}$.  The overplotted curves show the theoretically expected distribution for the expected number of degrees of freedom, $N=29$, using eq.\,\ref{pNoise}.  The red curve shows the expected distribution, while the green curve shows the distribution that would be derived if the S/N values derived by RM synthesis were 8.5\% higher than expected (see text for discussion).}\label{fig:simsPamp}
\end{figure}

For each source the ``signal'' was extracted at the value of $\phi$ at which the amplitude in the Faraday dispersion spectrum was a maximum.  The statistics of the polarization amplitude at a fixed Faraday depth, $\phi=0$, were also recorded.  The resulting probability distributions are shown in Fig.\,\ref{fig:simsPamp}.  There is excellent agreement between the distribution for the $\phi=0$ case and the Rayleigh distribution discussed in \S\ref{subsubsec:ConstPhi}.  The lower panel of Fig.\,\ref{fig:simsPamp} shows the amplitude distribution determined from the synthetic noise dataset when the RM synthesis algorithm is allowed to find a source anywhere over the Faraday depth range, $[-4000,+4000]$\,rad\,m$^{-2}$.  One expects na\"ively, on the basis of the width of the RMTF and the range of the search in Faraday depth, that there are $N\approx 29$ independent Faraday depth trials over the search space.  There is good agreement in the shape of the theoretically- and empirically-derived distributions for this value of $N$.  However, it is evident that the theoretical distribution peaks at a lower S/N than the empirically-derived distribution.  This cannot be accounted in terms of a change in $N$ since, although this would shift the peak of the distribution to higher S/N, it would also narrow the distribution to a value lower than is observed.  We note that, from a purely empirical viewpoint, the two distributions agree closely if the S/N axis is scaled such that the recovered S/N values from the simulated sources are systematically higher than those expected in the theoretical distribution by 8.5\%. The distribution corresponding to this is also shown in Fig.\,\ref{fig:simsPamp}.  However, the reason for such a S/N discrepancy is not obvious; the lower right panel of Fig.\,\ref{fig:brute force} shows that there is no systematic bias in the S/N of sources recovered by the RM synthesis technique and a brute-force fit to the polarization data.

\begin{figure}[ht]
\begin{center}
\includegraphics[angle=0,width=0.85\textwidth]{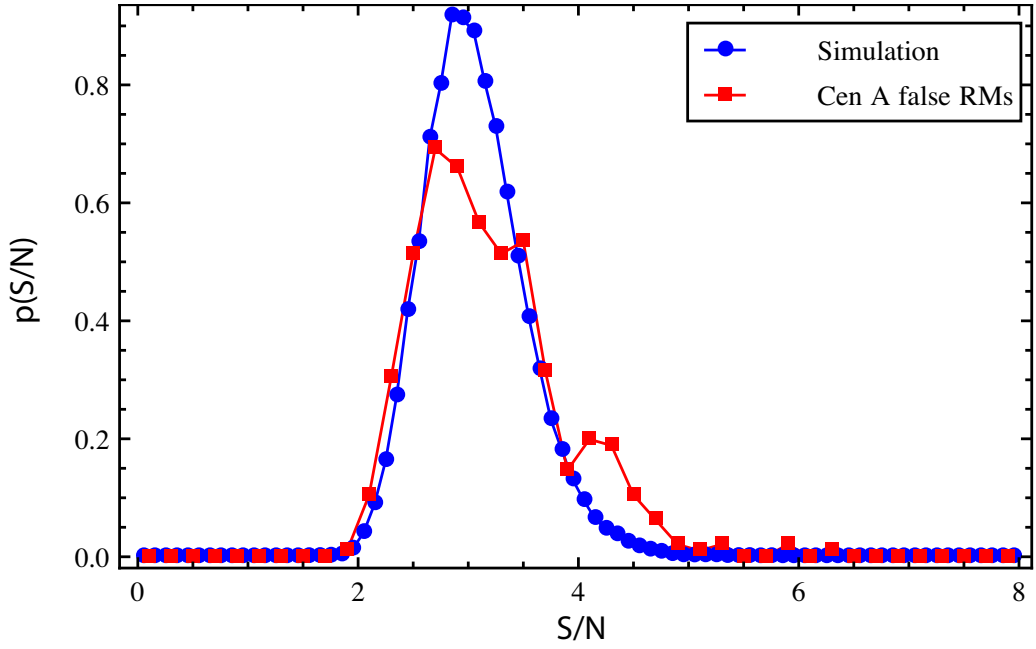} 
\end{center}
\caption{A comparison of the probability distribution of polarization amplitudes found in the Cen A data which are regarded as having ``false'' RM identifications with the distribution expected of a purely noise-like signal based on a simulation of $5 \times 10^4$ synthetic sources.  The error bars on the Cen A source counts reflect the fact that the distribution is estimated from a finite (and small) number of sources, and scale as the square-root of the number of sources per bin.  
}\label{fig:simsCompare}
\end{figure}

One can directly compare the polarization amplitude distribution for the sources in the Cen A field with ``false'' RMs (see \S\ref{subsec:StoN}) against the distribution, $p_n$ as determined by the synthetic pure-noise sources.  This comparison is shown in Fig.\,\ref{fig:simsCompare}. 
In this plot we see excellent agreement between the polarization amplitude distribution of sources with false RMs and the expected noise distribution.  However, there is an excess of sources in the region 4$<$S/N$<$5 which deserves further investigation.  A likely explanation for this excess is that a number of these sources are mistakenly identified as false detections.  It is possible that several contain real signal, but their S/N is sufficiently low that their Faraday depth is poorly localized by RM synthesis (i.e.\,they are only localized to about the $\sim 280\,$rad\,m$^{-2}$ width of the RM transfer function), thus placing the detection RM outside the nominal range $[-210,110]$\,rad\,m$^{-2}$ considered to constitute a ``good'' detection.  The excess of sources can be seen in Fig.\,\ref{fig:SNhistogram} as the concentration of sources in the $4<{\rm S/N}<5$ bin with RMs in the range $[0, 500]\,$rad\,m$^{-2}$. 
The results of \S\ref{subsec:sigPlusnoise} (and Fig.\,\ref{fig:P0alldistn}) indicate that when some small signal is present the polarization amplitude distribution shifts to higher S/N.  This suggests that some sources with supposedly ``false'' RMs do indeed contain intrinsic signal.

\begin{figure}[ht]
\begin{center}
\includegraphics[angle=0,width=0.85\textwidth]{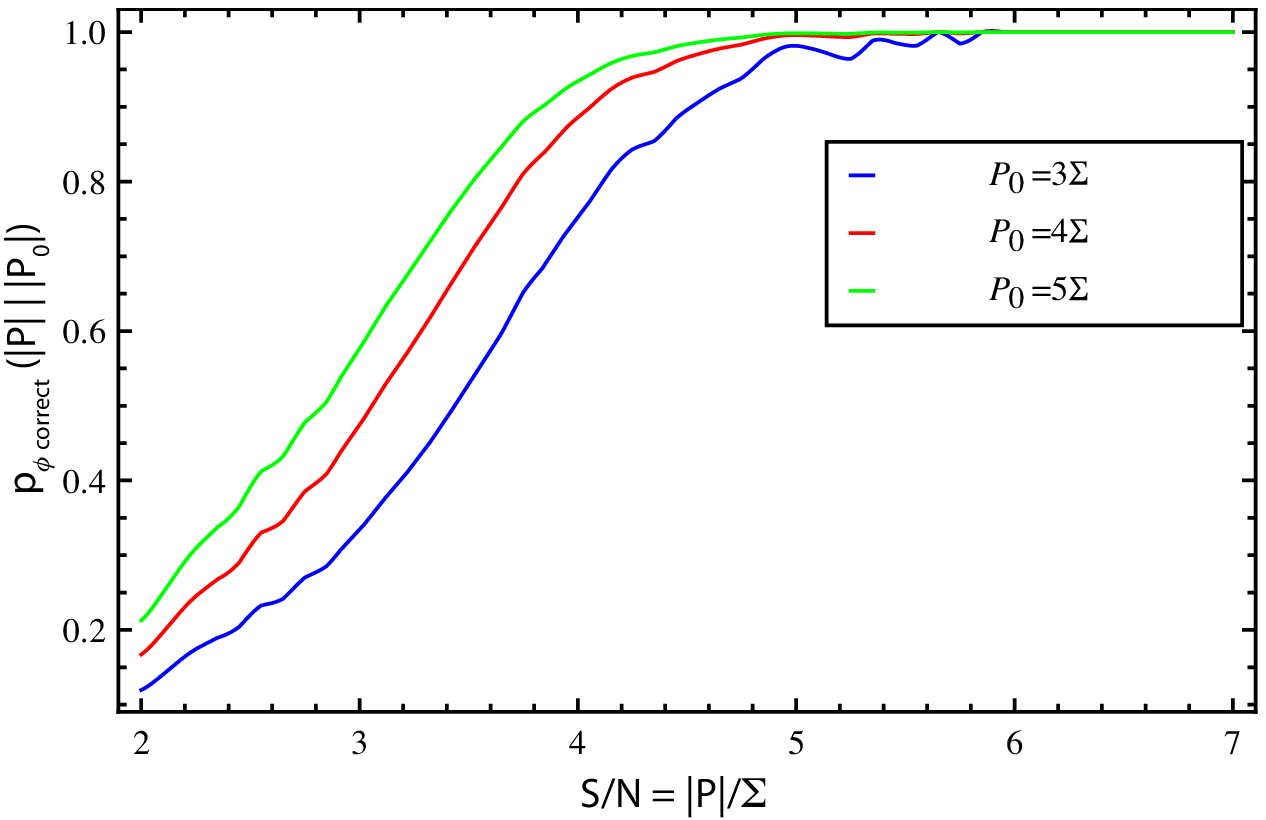} \\
\end{center}
\caption{The probability that RM synthesis correctly identifies the Faraday depth of the polarized signal as a function of the measured polarization amplitude, $P$.  The curve is calculated for intrinsic signals of $|{\cal P}_0|/\Sigma = 3, 4$ and $5$.  The small oscillations in the curves are due to the fact that the noise distribution, $p_c$, is not computed from an analytic expression, but from the histogram derived from the simulation of $5\times 10^4$ purely noise-like sources.} \label{fig:Pphicorrect}
\end{figure}

A related statistic is $p_{\phi \hbox{ correct}}$, the expected fraction of sources whose Faraday depths are identified correctly as a function of $P$ for a given $P_0$.  We compute this quantity using the numerically-derived distribution, $p_c$.  Figure \ref{fig:Pphicorrect} displays this likelihood.  This gives a direct estimate that the probability is correct at any given RM.   At S/N $\approx 3$ the probability of retrieving the correct Faraday depth is $\approx 50$\%, and it approaches 100\% for S/N values between 5 and 6, depending on the value of the intrinsic polarization value, $P_0$,

It is interesting to compare the differential source counts from the Cen A data in Table\,\ref{tab:detections} against this curve, despite uncertainty in the intrinsic polarization amplitudes that contribute to the Cen A data.  At S/N values of 3, 4 and 5, the measured fractions of good RMs are 26\%, 57\% and 92\% respectively, while the predicted fractions at these S/N values are in the ranges 25-57\%, 61-93\% and 94\%-100\% respectively. In estimating the range of fractions, we have assumed that the S/N of the intrinsic polarization, $P_0$, is between 2.5 and 5.

\subsubsection{Forward Modelling of Polarization Source Counts}

It is beyond the scope of this paper to implement a complete iterative deconvolution scheme to recover the intrinsic polarization source counts, $N(P_0)$, using eq.(\ref{SrcCounts}) in conjunction with $p_c ( Z \,|\, P_0)$.  

\begin{figure}[ht]
\includegraphics[angle=0,width=0.9\textwidth]{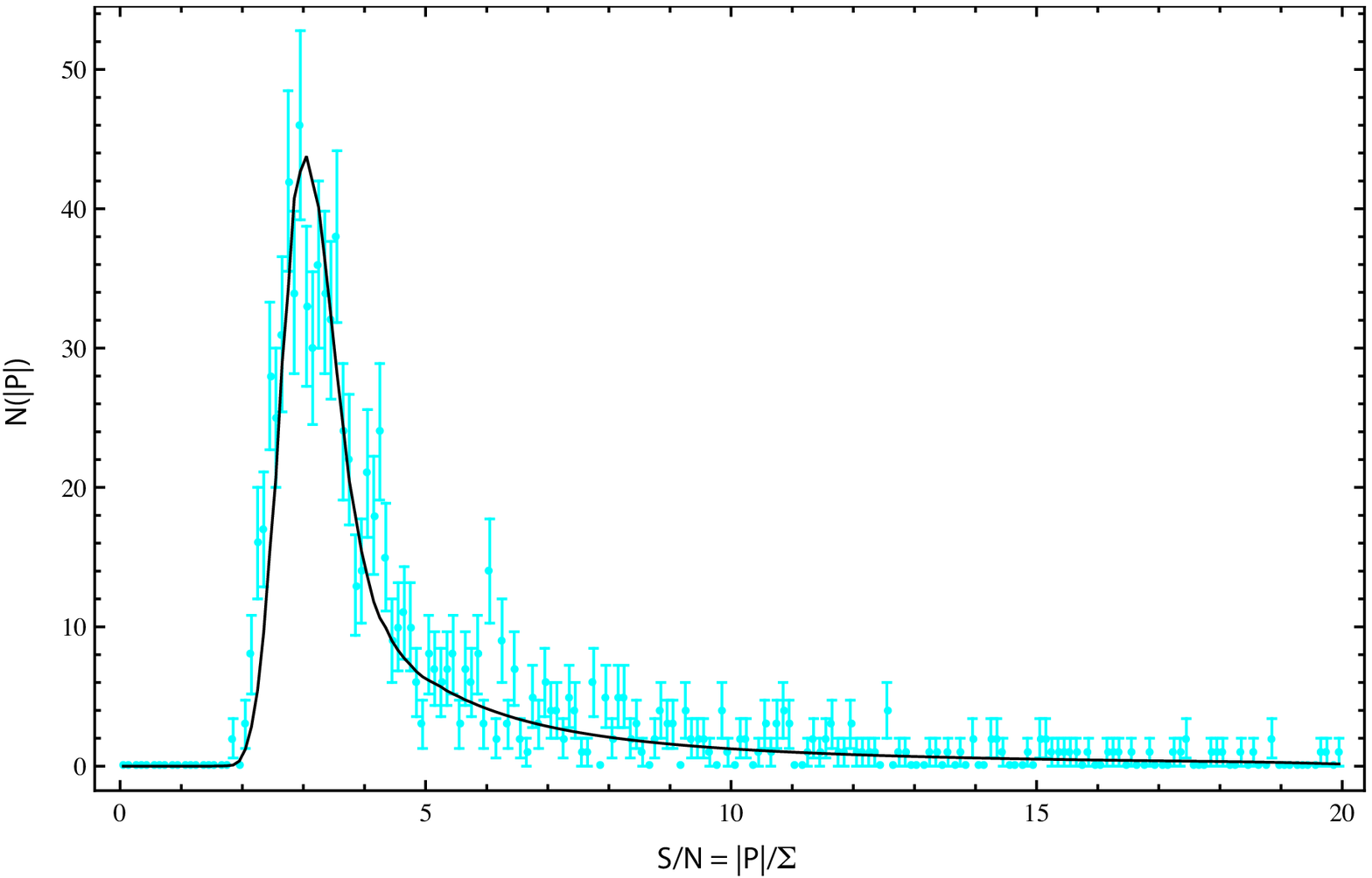}
\caption{The observed source counts in $P$ (cyan) and the best fitting model of it (black).}\label{fig:counts}
\end{figure}

We have instead developed a simple iterative forward-modelling scheme in which the intrinsic polarized source count distribution is taken to be of the form,
\begin{eqnarray}
N(  P_0  ) = A \left\{ \begin{array}{cc}
(C/\Sigma)^{\alpha_1} ( P_0 / C  )^{\alpha_2} , 	&  P_0 / \Sigma < C \\
( P_0 / \Sigma )^{\alpha_1},  & P_0 / \Sigma < C , \\
\end{array}
\right.
\end{eqnarray}
which is a broken power law distribution with index $\alpha_1 < 0$ that turns over at a S/N of $C$ and scales with an index $-4 < \alpha_2 < 4$ below this S/N break, and where $A$ is an overall normalization\footnote{Since the distribution $p ( P\, |\, P_0)$ is zero for $P_0/\Sigma \leq 2$, the fact that $N ( P_0 )$ diverges for small $P_0$ when $\alpha_2 < 0$ does not pose a problem for our modelling.  We are effectively insensitive to the behaviour of the distribution at $P_0/\Sigma \leq 2$.}.   This distribution was gridded onto the vector ${\bf y}$ with a resolution in $\Delta P/\Sigma = 0.1$ and substituted into eq.\,(\ref{SrcCountsMatrix}) to find the predicted distribution of measured $P$ values, which we label ${\bf x}_{\rm pred}$.  This is compared against the actual distribution vector ${\bf x}_{\rm meas}$ by computing the reduced-$\chi^2$ of the fit,
\begin{eqnarray}
\chi^2=\frac{1}{N_{\rm bins}-4} \sum_i^{N_{\rm bins}} \frac{|{\bf x}_{{\rm pred}_i} - {\bf x}_{{\rm meas}_i} |^2}{\sigma_i^2}.
\end{eqnarray} 
where we take the errors to be $\sigma_i^2 = x_{{\rm meas}_i}$ (and set $\sigma_i=1$ when $x_{{\rm meas}_i}=0$).  The quantity $\chi^2$ was searched for a minimum in terms of the parameters $A$, $C$, $\alpha_1$ and $\alpha_2$.  The minimum occurs at $\chi^2 = 0.85$ for $(C, \alpha_1,\alpha_2,A) = (3.72, -2.17,-0.30, 2.60 \times 10^4)$, and the best-fitting solution is plotted over the measured polarization distribution in Figure \ref{fig:counts}.  However, the minimum formally supports a range of indices in the range $-2.9 \lesssim \alpha_1 \lesssim -1.0$, and we note that the values of $\alpha_1$ and $\alpha_2$ and $C$ are all highly correlated.   To illustrate this, Table \ref{tab:fitres} below lists the derived fit parameters for a range in the break S/N, $C$.

\begin{table*}[ht]
\begin{center}
\begin{tabular}{| lcccc |}
\hline
$C$ & $\chi^2$  & $\alpha_1$ & $\alpha_2$  & $A$ \\
\hline
0.0 & 1.12 & -0.96 & -- & 119 \\
1.0 & 0.93 & -1.53 & 4 & 1.56 $\times 10^3$ \\
2.0 & 0.88 & -1.90 & 1.01 & 8.09 $\times 10^3$ \\
3.0 & 0.86	& -2.04 & -0.05 & 1.48 $\times 10^4$ \\
3.72 & 0.85 & -2.17 & -0.30 & 2.60 $\times 10^4$ \\
4.0 & 0.85 & -2.22 & -0.37 & 3.27 $\times 10^4 $\\
5.0 & 0.87 & -2.40 & -0.54 & 7.39 $\times 10^4$ \\
6.0 & 0.91 & -2.54 & -0.66 & 1.45 $\times 10^5$ \\
7.0 & 0.95 & -2.64 & -0.74 & 2.50 $\times 10^5$ \\
\hline
\end{tabular}
\end{center}
\caption{The fit quality and associated fit parameters as a function of location in the spectral break, $C$. This table illustrates the covariance between the break S/N and the spectral indices of the polarization amplitude count distribution.} \label{tab:fitres}
\end{table*}

%Figure \ref{fig:counts} shows the source counts in both $I$ and $| {\cal P}|$. 
The best-fitting index of the slope of the differential counts in $P$ at high S/N, $\alpha_1=-2.17$, is very close to the value of $-2.2$ obtained by fitting just the stronger well-detected sources in the sample with $P$ in the range 0.6 to 6\,mJy.  This indicates that the source counts in $P$ continue with the same slope down to at 0.35\,mJy, and possibly lower.  Note that this forward modelling  method does not require any assumptions about polarization amplitude bias at low S/N.

Following the technique used by Tucci et al.\,(2004) we can now estimate the change in fractional polarization as a function of flux density by comparing the source counts in $I$ and $P$ at the same source density.  For both FIRST (White et al.\,1997) and these Centaurus A field sources the differential counts in $I$ have a slope of $-1.8$ in this flux range.  The fractional polarization is 3.0\% for the strongest sources ($>200\,$mJy) and this increases to 3.5\% for sources below 20 mJy.  This can be compared with the mean fractional polarization of ATLSB sources ($S_\nu >$ 0.4\,mJy) of 4.3\%.  Taylor et al.\,(2007) make a detailed analysis of polarization in the Elais N1 field.  They have 786 sources with 83 polarization detections.  We have 1005 sources and at least 346 polarization detections $(5 \Sigma)$ in the same flux density range.  We agree well in the source density and polarized fraction at the higher flux density but our fractional polarization for sources below 20\,mJy (3.7\%) is less than the increase to 4.8\% seen by Taylor et al.\,(2007).

\section{Sources with complex polarization} \label{sec:complex}
 
To evaluate the reality of sources which have multiple RMs we evaluated the reduced $\chi^{2}$ after RM synthesis identified each new CLEAN component\footnote{Recall that CLEAN recognised all  components that were within 5\,rad\,m$^{-2}$ of each other as being part of the same component.}, and used this to help determine the significance of each new component.  To evaluate the success of this process and to see what type of source has high reduced $\chi^{2}$ we looked at the 21 sources with reduced $\chi^{2} > 2.4$ and S/N $>$ 5.  We can divide these sources into four categories:
 
\begin{enumerate}
\item Some residual bad data points were readily identified in this process and these (6) sources do not appear in the list when the bad data is flagged out.
\item High S/N (3 sources).  The high $\chi^{2}$ associated with some high S/N sources could be due to real but weak components with different RM which are only visible at high S/N.   It may also result from instrumental dynamic range errors.
\item Multiple RM (8 sources).  These are well fit by multiple components.  Note that for sources with multiple RMs we are unable to uniquely separate components within our RM resolution of $\sim 280\,$rad\,m$^{-2}$.
\item Significant spectral change in $P$ across the band (4 sources).  This is likely to be caused by regions with closely spaced RM and polarization vectors as discussed by Farnsworth et al.\,(2011), although it may in principle be caused by depolarization effects.
\end {enumerate}

\begin{figure}[ht]
\begin{center}
\begin{tabular}{c}
\includegraphics[angle=0,width=0.45\textwidth]{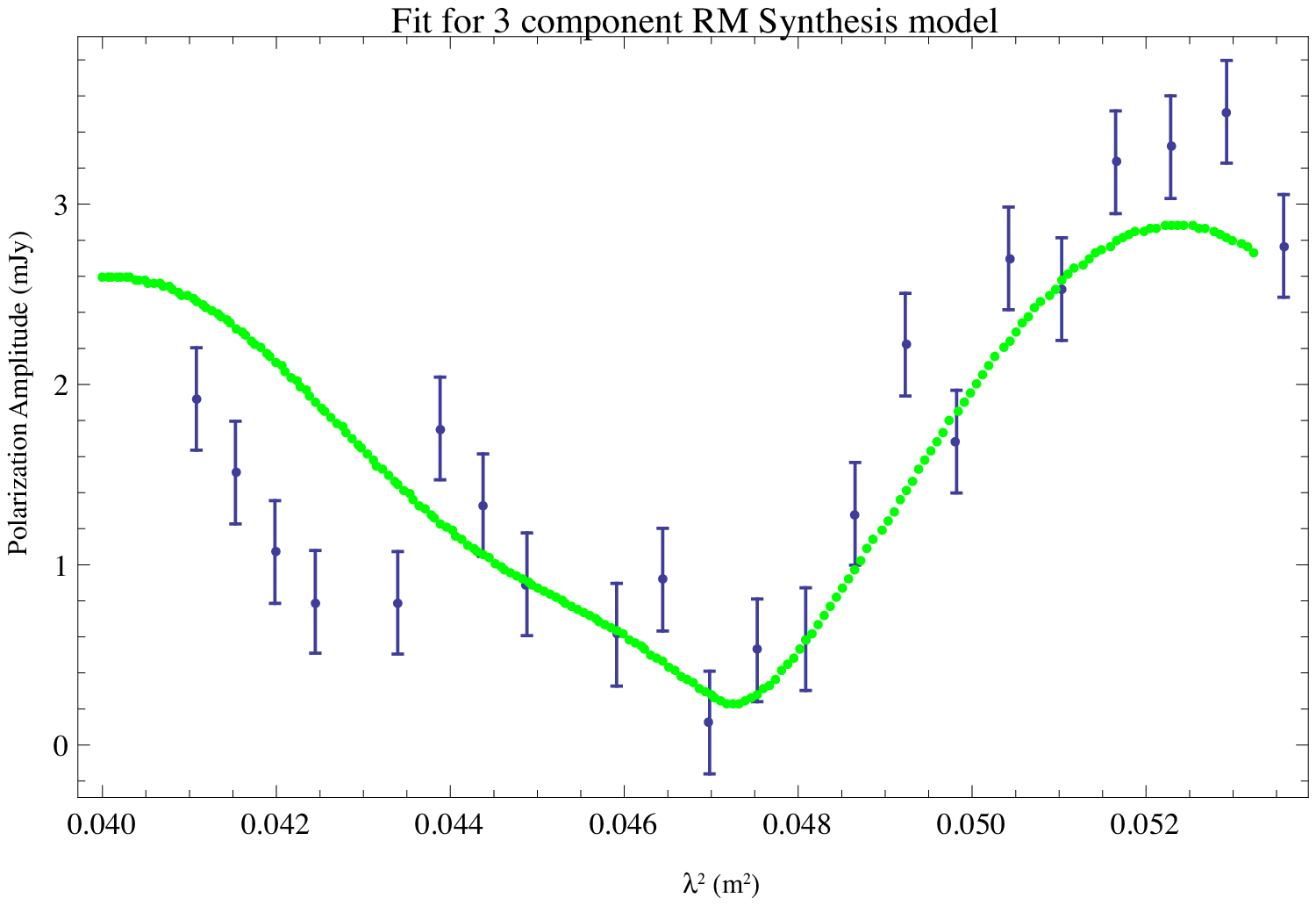} 
\includegraphics[angle=0,width=0.45\textwidth]{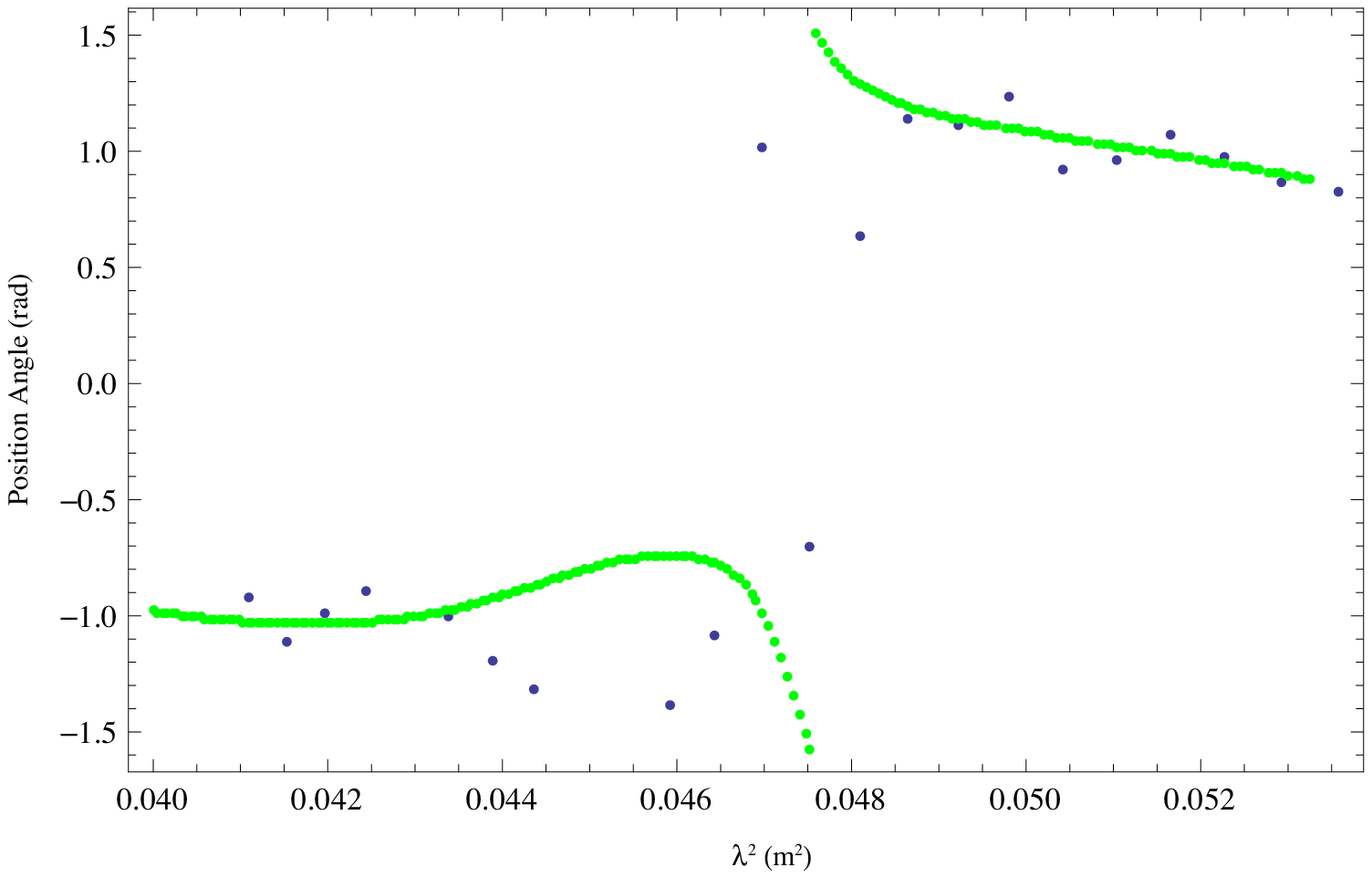} \\
\includegraphics[angle=0,width=0.55\textwidth]{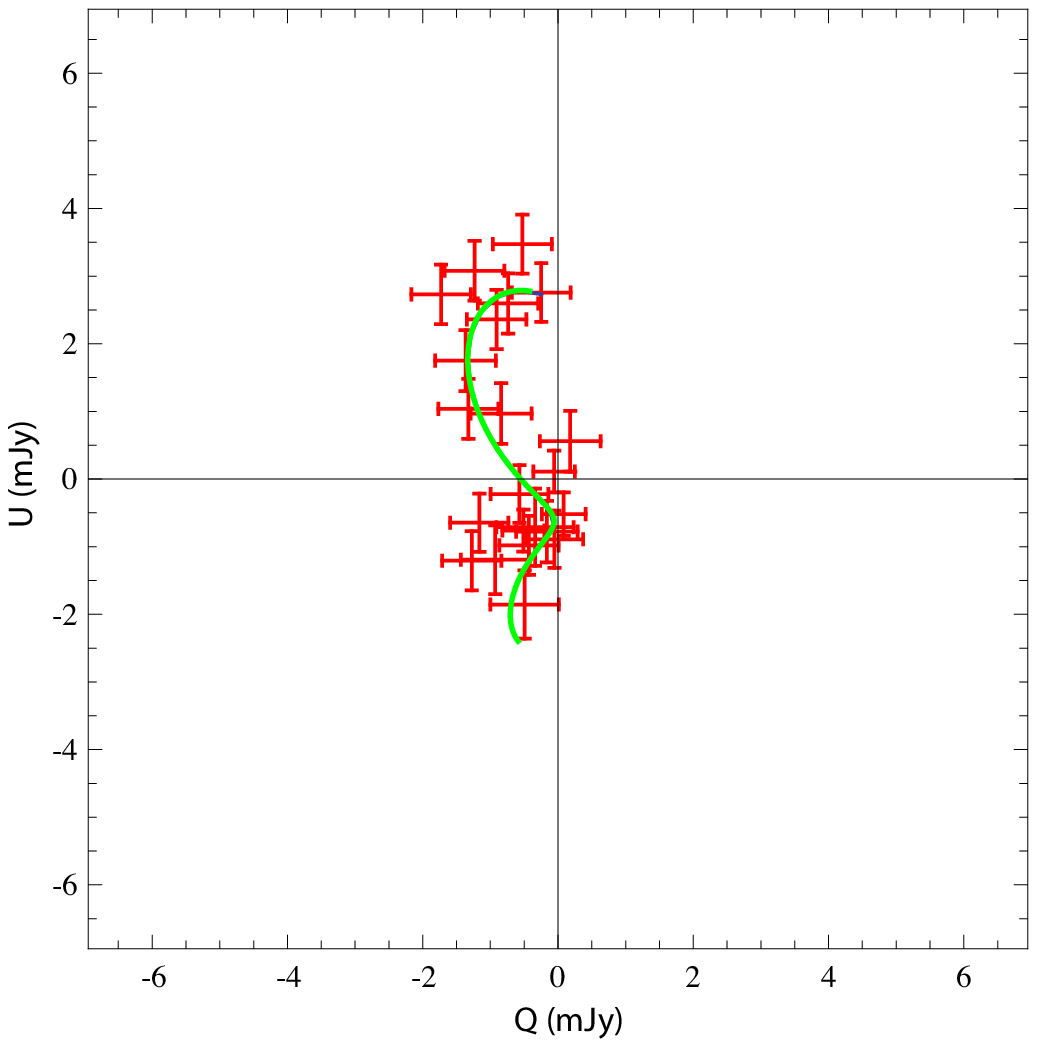} \\
\end{tabular}
\end{center}
\caption{The 3-component polarization solution for the source identified in Feain et al.\,(2009) as 131943$-$445904. The fit parameters for each component are shown in Table \ref{tab:Eg1}, including the reduced $\chi^2$ value after each successive component is added to the model.}\label{fig:ComplexEg1}
\end{figure}

\begin{table}[ht]
\caption{The fit parameters for the source 131943$-$445904, as shown in Fig.\,\ref{fig:ComplexEg1}.} \label{tab:Eg1} 
\begin{tabular}{lcccc}
\hline
Component no. & RM (rad\,m$^{-2}$) & reduced $\chi^2$ after inclusion & $ P_0$ (mJy) &  S/N \\
\hline
1      &        -120.4    & 	4.5 &  	1.41     &                            		15.0  \\
2      &       171.2       &  2.6 & 		0.94     &                           10.0 \\
3      &      -323.2      & 1.6 & 		0.54        &                        5.8 \\
\hline
\end{tabular} 
\end{table}

\begin{figure}[ht]
\begin{center}
\begin{tabular}{c}
\includegraphics[angle=0,width=0.45\textwidth]{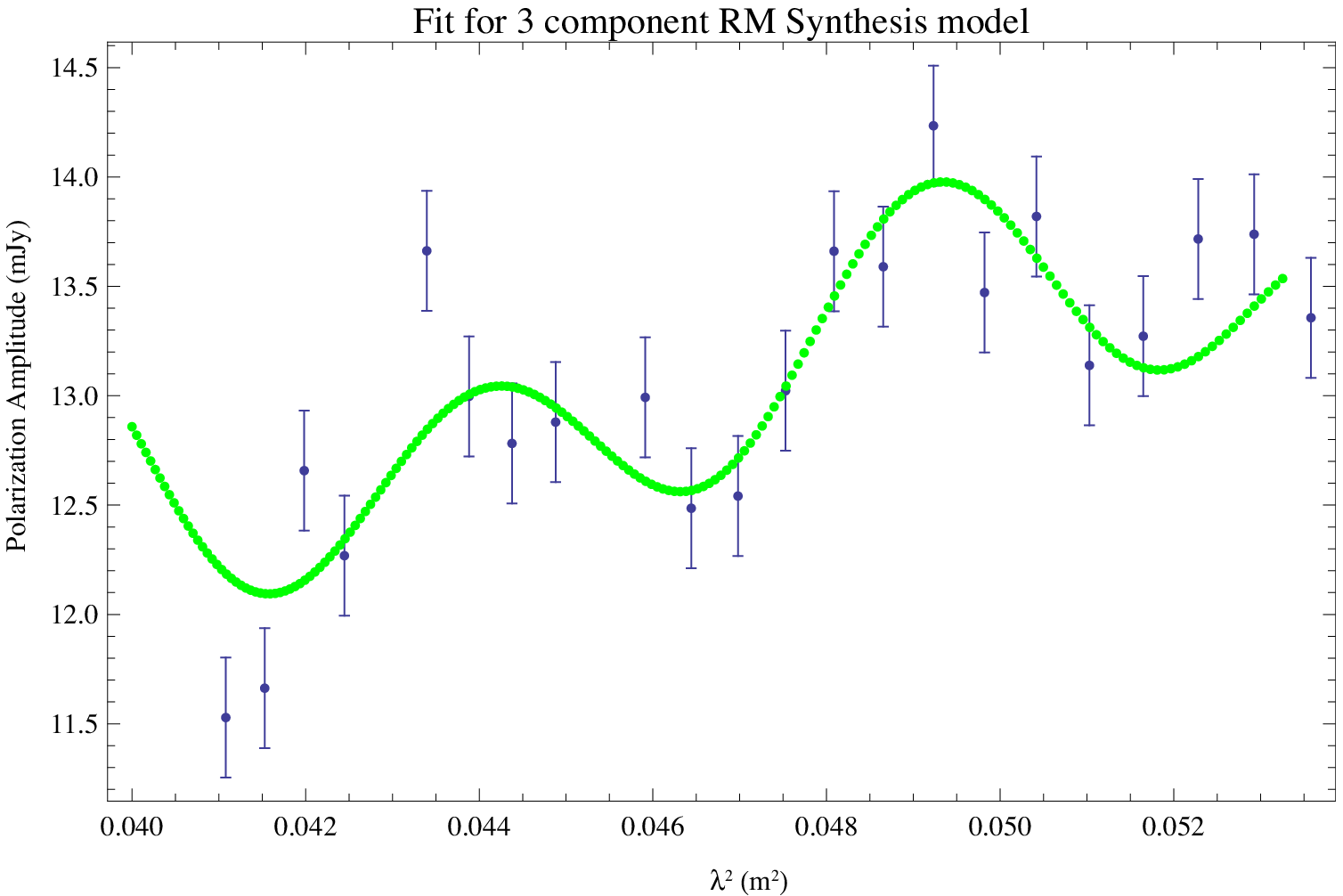} 
\includegraphics[angle=0,width=0.45\textwidth]{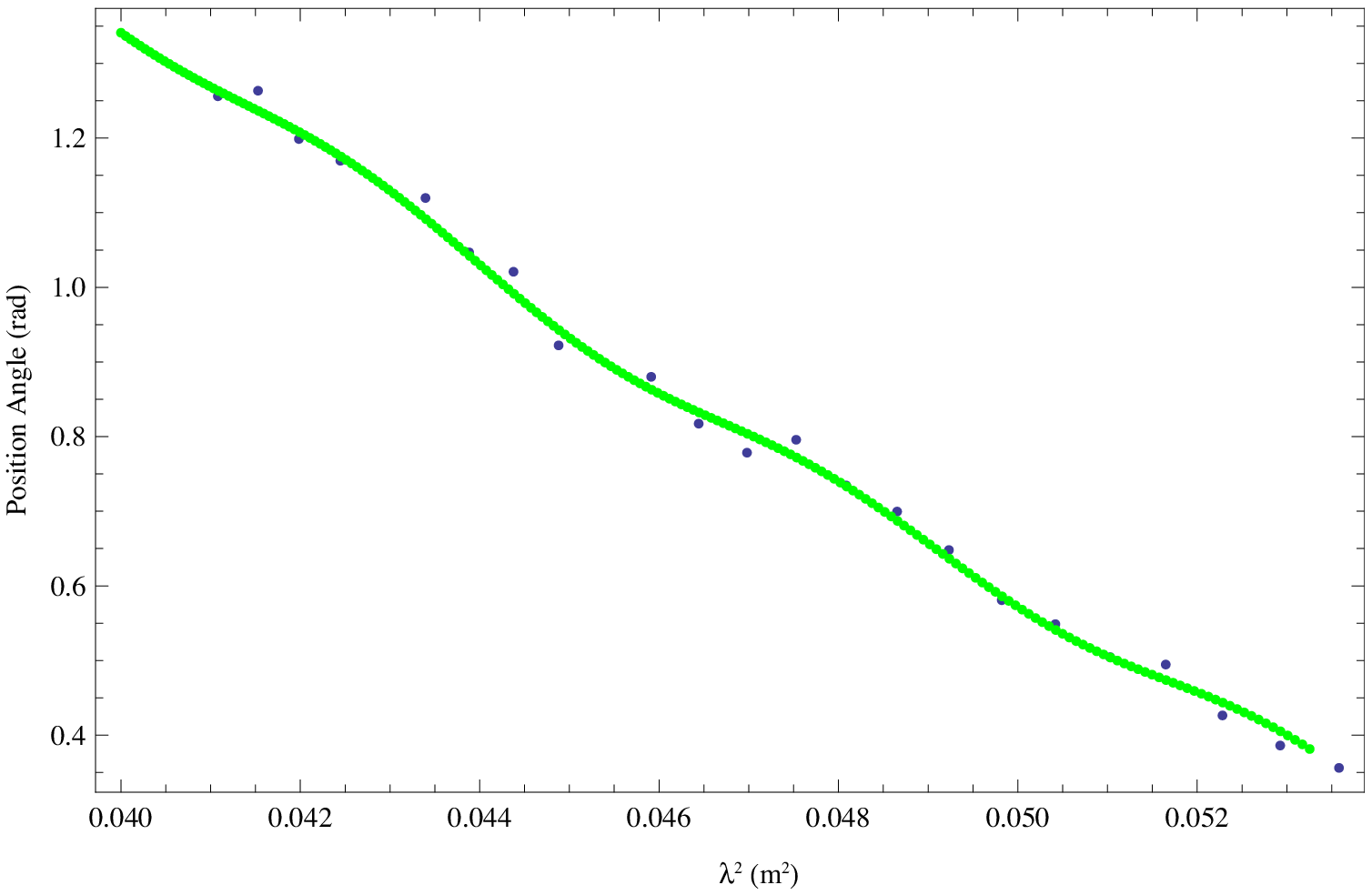} \\
\includegraphics[angle=0,width=0.55\textwidth]{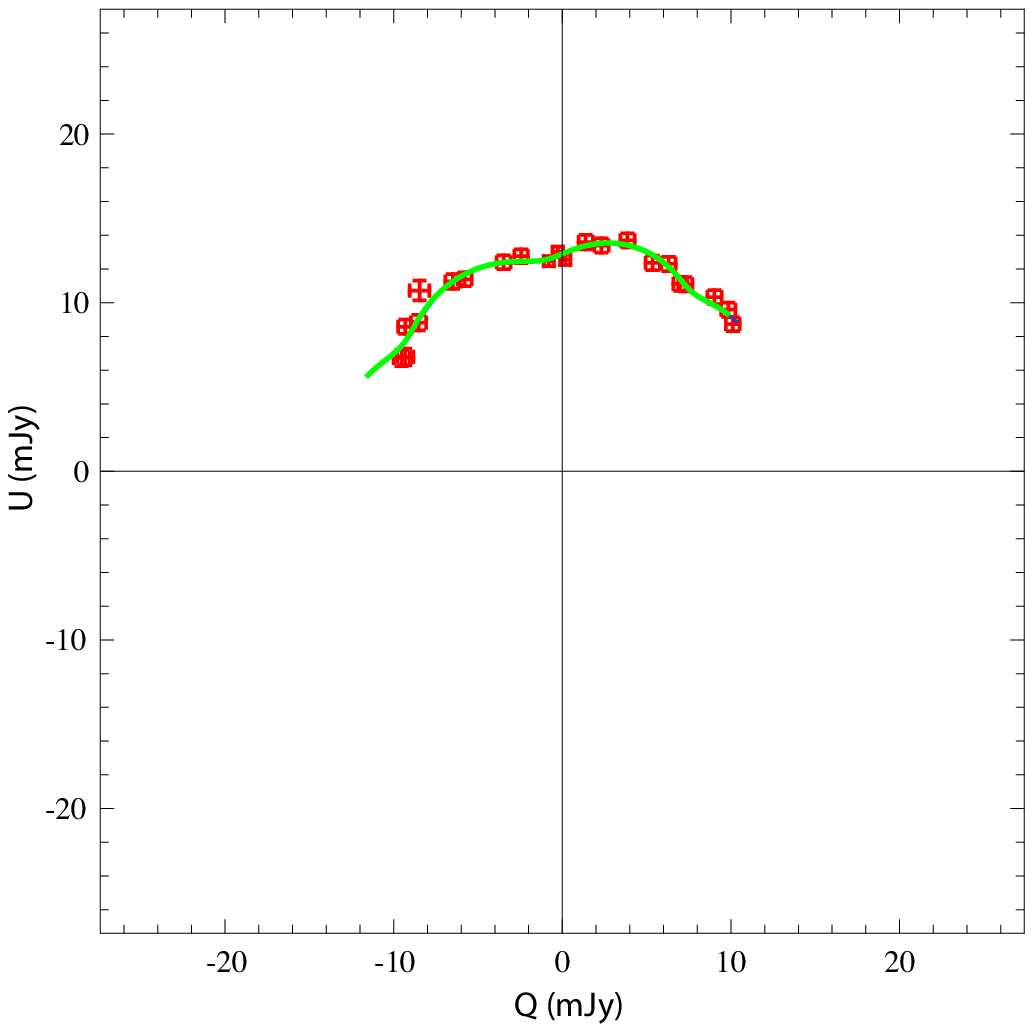} \\
\end{tabular}
\end{center}
\caption{The 3-component polarization solution for the source identified in Feain et al.\,(2009) as 131713$-$410934.  The fit parameters for each component are shown in Table \ref{tab:Eg2}, including the reduced $\chi^2$ value after each successive component is added to the model. 
}\label{fig:ComplexEg2}
\end{figure}

\begin{table}[ht]
\caption{The fit parameters for the source 131713$-$410934, as shown in Fig.\,\ref{fig:ComplexEg2}.} \label{tab:Eg2}
\begin{tabular}{lcccc}
\hline
Component no. & RM (rad\,m$^{-2}$) & reduced $\chi^2$ after inclusion & $P_0 $ (mJy) &  S/N \\
\hline
1      &        -74.8    & 	2.7 &  	13.1     &                            		66.8  \\
2      &       118.4       &  1.7 & 		0.56     &                           2.9 \\
3      &      -686.8      & 1.1 & 		0.45        &                        2.3 \\
\hline
\end{tabular}
\end{table}
 
We illustrate these effects with two sources from this list:
 
131943-445904 has three significant and comparable amplitude RM components which are just separated at our RM resolution.  The three components are a good fit and reduce $\chi^{2}$ from 4.5 to 1.6.  See Figure\,\ref{fig:ComplexEg1} and Table \ref{tab:Eg1}.
 
131713-410934 is a high S/N case with a dominant RM components at $-75\,$rad\,m$^{-2}$.  It also has a barely resolved component at $118$\,rad\,m$^{-2}$ which causes the slope in $P$ and a much weaker (3\%) well resolved component at RM$=-687$\,rad\,m$^{-2}$.  See Figure\,\ref{fig:ComplexEg2} and Table \ref{tab:Eg2}.
 
It is beyond the scope of this paper to present a detailed discussion of the sources with complex RM structure but we can make a few useful observations.  Sources with RM structure less than our resolution (280 rad\,m$^{-2}$) can still cause strong variations in polarization amplitude across the band and the RM clean process can fit this change in amplitude with components separated by approximately the HPBW.   While this is still a valid indication of significant RM structure, the ``super resolution'' of our CLEAN process (i.e. the ability to resolve RM components within the width of the RMTF) is unlikely to find a unique solution for the actual RM components.

Although we do see clear evidence of multiple RMs in 12 of the 359 sources with S/N in polarization amplitude greater than 5, this is only 3\% of our sample and the majority of sources are well fit by a single RM component.  

To better quantify this fraction we can ask down to what S/N one can we detect a second RM component by examining to what degree the addition of the second component improves the fit to the spectropolarimetric data.  We start with a model consisting of one polarized component ${\bf M}_1$ and compare it against the measurement vector ${\bf m}$.  Let us assume that the measurements in each spectral channel have the same error, $\sigma$.  Then, if the vector ${\bf m}$ contains $N_{\rm chan}$ components, one has
\begin{eqnarray}
\chi_1^2= \frac{1}{ (2N_{\rm chan} - 3) \sigma^2} ({\bf M}_1 - {\bf m})^2. 
\end{eqnarray}
Now let us introduce a second polarized component, ${\bf M}_2$ that improves the fit and produces a new reduced $\chi^2$ of 
\begin{eqnarray}
\chi_2^2 = \frac{1}{(2N_{\rm chan} - 6) \sigma^2 } ({\bf M}_1+{\bf M}_2 - {\bf m})^2.
\end{eqnarray}
Subtracting the two foregoing expressions and, if the data are well explained by two polarized components, one has ${\bf M}_1 - {\bf m} \approx {\bf M}_2$ we find, 
\begin{eqnarray}
\Sigma^2 [\chi_2^2 (2N_{\rm chan}-6) - \chi_1^2 (2N_{\rm chan}-3)] \approx 3 |{\bf M}_2|^2 / N_{\rm chan},
\end{eqnarray}
where we use the fact that $\Sigma = \sigma/\sqrt{N_{\rm chan}}$.  Recognising that $|{\bf M}_2|^2/N_{\rm chan}$ is the average squared amplitude of the second polarized component, say $ P_2^2$, we find,
 \begin{eqnarray}
\frac{ P_2^2}{\Sigma^2} \approx \frac{1}{3} [\chi_2^2 (2N_{\rm chan}-6) - \chi_1^2 (2N_{\rm chan}-3)]
\end{eqnarray}
For our measurements, one has $N_{\rm chan} = 22$ and we can see that, to improve the fit from $\chi_1^2 =2.4$ to $1.3$, one reqires a component with a S/N of $P_2/\Sigma \approx 3.6$.

We note that the statistical treatment for the detection of a second RM component in the presence of another strongly polarised signal is very different from the analysis of the detection of a polarised signal in the presence of noise.  The perturbation that an additional polarised signal generates will be closer to the normal detection statistics so the weak multiple components we find at the S/N $\gtrsim 3$ level are already significant.

\section{Conclusions}\label{sec:conclude}

Our major conclusions are as follows:

\begin{itemize}
\item We have assessed the distribution of false and correct RM detections in the 1005-source sample of Feain et al.\,(2009).  This yields a quantitative estimate of the likelihood that a given RM, as found by RM synthesis, is correct, as a function of S/N.   If $4<$S/N$<5$, we find the likelihood of finding a correct RM is about 52\%.  Restricting the range of possible RM solutions to the range -210 to 110\,rad\,m$^{-2}$ (the range found for the high S/N sources) further decreases the likelihood of a false RM detection; there is only a 4\% chance that any given Faraday depth found by RM synthesis in a search over the range $[-4000,4000]$\,rad\,m$^{-2}$ will fall in the range of expected real detections by accident.
\item There is no systematic difference in the polarization amplitude recovered by RM synthesis compared to a least-squares fitting approach as a function of S/N, and the average ratio of the amplitudes found by the two methods is $1.0$.  
\item We have investigated the effect of noise in RM synthesis and examined the probability distribution of the polarization amplitude as determined by this algorithm.  From this, we have developed a formalism to recover the distribution of intrinsic polarization amplitudes from a measured distribution which is affected by noise.  

This enables us to recover the polarization distribution at S/N levels well below the confidence level for any single detection.  We have applied this to examine the polarization counts of the sources in the Feain et al.\,(2009) Cen A catalog; the best fitting polarization source count distribution follows $N \propto P^{-2.35}$, which is consistent with $N \propto P^{-2.16}$ found by fitting only to S/N$>7$ detections, and with no turnover at an amplitude above 0.2\,mJy.  

A related point is that polarization ``bias'' should not be subtracted from the derived polarization amplitudes when analysing the distribution of $P$.  Subtraction of this bias is often at best misleading, since the distribution of $P$ is highly non-Gaussian.  
\item Pre-existing analyses have used a S/N cutoff of $\approx 7$ as a limit to the believability of results from RM synthesis.  We derive here the likelihood that a given RM detection is correct as a function of S/N and the intrinsic polarization amplitude ($P_0$) of the polarization signal.  We also derive the criteria for the likelihood of the detection of a second RM component.  For instance, a detection at a S/N $\sim 3$ is sufficient to believe the existence of a second component if the addition of the second component reduces the $\chi^2$ value of the fit to the polarization data from $\approx 2.4$ to $\approx 1.3$.  
\end{itemize}

\acknowledgements
The Australia Telescope Compact Array is part of the Australia Telescope which is funded by the Commonwealth of Australia for operation as a National Facility managed by CSIRO.

{\it Facilities:} \facility{ATCA}

\end{document}